\documentclass[12pt,english,aps,manuscript]{revtex4}
\usepackage[T1]{fontenc}
\usepackage[latin9]{inputenc}
\usepackage{units}
\usepackage{amsmath}
\usepackage{amssymb}
\usepackage{graphicx}
\usepackage{esint}

\makeatletter

\providecommand{\tabularnewline}{\\}

\@ifundefined{textcolor}{}
{%
 \definecolor{BLACK}{gray}{0}
 \definecolor{WHITE}{gray}{1}
 \definecolor{RED}{rgb}{1,0,0}
 \definecolor{GREEN}{rgb}{0,1,0}
 \definecolor{BLUE}{rgb}{0,0,1}
 \definecolor{CYAN}{cmyk}{1,0,0,0}
 \definecolor{MAGENTA}{cmyk}{0,1,0,0}
 \definecolor{YELLOW}{cmyk}{0,0,1,0}
}

\makeatother

\usepackage{babel}
\begin{document}

\title{Three-dimensional Simulation of Thermal Harmonic Lasing FEL with
Detuning of the Fundamental}

\author{E. Salehi$^{(1)}$, B. Maraghechi$^{(1,2)}$ }

\email{behrouz@aut.ac.ir}

\author{, N. S. Mirian$^{(2,3)}$ }

\address{$^{(1)}$Department of Physics, Amirkabir University of Technology,
Postal code 15875-4413, Tehran, Iran.}

\address{$^{(2)}$School of Particle and Accelerator Physics, Institute for
Research in Fundamental Sciences (IPM), Postal code: 19395-5531, Tehran,
Iran}

\address{$^{(3)}$UVSOR Facility (UVSOR) Institute for Molecular Science,
444-8585, Myodaiji J - Okazaki, japan.}

\begin{abstract}
Detuning of the fundamental is a way to enhance harmonic generation
. By this method, the wiggler is composed of two segments in such
a way that the fundamental resonance of the second segment to coincide
with the third harmonic of the first segment of the wiggler to generate
extreme ultraviolet radiation and x-ray emission. A set of coupled,
nonlinear, and first-order differential equations in three dimensions
describing the evolution of the electron trajectories and the radiation
field with warm beam is solved numerically by CYRUS 3D code in the
steady-state for two models (1) seeded free electron laser (FEL) and
(2) shot noise on the electron beam (self-amplified spontaneous emission
FEL). Thermal effects in the form of longitudinal velocity spread
is considered. Three-dimensional simulation describes self-consistently
the longitudinal spatial dependence of radiation waists, curvatures,
and amplitudes together with the evaluation of the electron beam.
The evolutions of the transverse modes are investigated for the fundamental
resonance and the third harmonic. Also, the effective modes of the
third harmonic are studied. In this paper, we found that detuning
of the fundamental with shot noise gives more optimistic result than
the seeded FEL.
\end{abstract}
\maketitle

\section{Introduction}

Fourth generation light sources have given us the capability of exploring
the molecular and atomic phenomena. High-gain free electron laser
(FEL) amplifiers hold great prospects of reaching coherent high power
radiation in the x-ray region of the electromagnetic spectrum. In
recent years, a great effort of researchers has been devoted to study
the process of higher harmonic generation in achieving lasing at shorter
wavelengths. A possible way for obtaining x-ray wavelength is by using
nonlinear harmonic generation \cite{ackerman,Bonifacio,Freund2000,Huang-1,saldin2006,colson}.

Radiation of the electron beam in the planar wiggler contains odd
harmonics but the output power at the $h$th harmonic is rather small
and is of the order of $10^{-h}$ times the power of the fundamental
\cite{ackerman,Freund2000,Huang-1,saldin2006,Dattoli1}. Various schemes
have been proposed to enhance the harmonic intensity, these are high-gain
harmonic generation\cite{yu1,yu2,Babzien,Giannessi,labat}, echo-enabled
harmonic generation using the beam echo effect \cite{stupakov1,stupakov2},
and the two-beam FEL for frequency up-conversion \cite{mcneil1,rouhani,mirian}.
Another technique is two harmonic undulators in which an on axis field
oscillates with integer multiples of spatial periods and with different
orthogonal polarizations \cite{dottoli2002}. The analysis of this
method is achieved with PROMETEO code developed by Dattoli and coworkers
\cite{prometeo}, which has been benchmarked with MEDUSA. Recently,
McNeil et al. \cite{Mcneil2} proposed a harmonic lasing method for
FEL amplifiers that can amplify the higher harmonics by suppressing
the interaction at the fundamental resonance. They showed that this
configuration can significantly extend the operation band of user
facilities.

Reference \cite{Mcneil2} has outlined two methods for suppressing
the interaction at the fundamental resonance while allowing the third
harmonic to evolve to saturation. The first method is based upon the
shifting of the phase of the fundamental between the wiggler segments,
which can be controlled by various techniques \cite{schneidmiller1}.
For the $h$th harmonic, this phase shift should be $2\pi n/h$, where
$n=1,2,3,\ldots$ is an integer number and $h=3,5,7,\ldots$ is the
harmonic number. Freund and coworkers \cite{freud2} investigated
the effects of varying the gap lengths between the wiggler segments
and varying the electron beam $\beta$ function on the third harmonic
generation with phase shift of $2\pi/3$. The second method is the
detuning of the fundamental by considering two different segments
for the wiggler. Two segments of the wiggler have different magnetic
field intensity while the wiggler period, $\lambda_{w}$, and the
initial average electron beam energy,$\gamma$, are kept constant.
It has been suggested in Ref. \cite{schneidmiller1} that a combined
use of the phase shifting and detuning of the fundamental can reduce
the bandwidth and increase the brilliance of the x-ray beam at saturation.
In Ref. \cite{Salehi}, detuning of the fundamental has been compared
for the nonaveraged and averaged methods in one-dimension and it is
shown that there is a remarkable agreement between the averaged and
nonaveraged simulations for the evolution of the third harmonic.

The method of detuning of fundamental has not been studied in three-dimensions
in the nonaveraged procedure. The 3D averaged wiggler method for detuning
of the fundamental method was studied in \cite{schneidmiller1}; they
showed that the gain length of detuning for the fundamental is larger
than that of the phase shift method (in the case of negligible energy
spread) and this result is consistent with the 1D theory \cite{Mcneil2}.
They also showed that 3D effects actually improve the situation and
makes harmonic lasing even more attractive compared to the 1D theory.
In this paper, three-dimensional features such as diffraction, radiation
guiding, and the evolution of transverse mode have been considered.
The guiding of laser light by an electron beam (optical guiding) in
an FEL occurs in the exponential gain regime \cite{moore2,moore,scharlemann}
when the coherent interaction between the source electron beam and
the electromagnetic field introduces an inward curvature in the phase
front of the light, which refracts it back toward the lasing core
of the electron beam. During the amplification, the electron beam
that propagates through the wiggler operates simultaneously as a guiding
structure that suppresses the diffraction, reducing the transverse
power losses, and enhancing the electromagnetic field amplification.
The evolution of transverse modes is important in planning for the
future user facilities that intend to employ radiation from this system
\cite{biedron}. 

The thermal effect of the electron beam is particularly important
for higher harmonics, because they are more sensitive to the energy
spread compared to the fundamental \cite{saldin2,yu3,Dattoli1}. The
energy spread is considered as a Gaussian energy distribution in MEDUSA
code for nonlinear harmonic generation \cite{Medusa3d}. Also, in
reference \cite{schneidmiller1}, the energy spread is considered
and the gain length for the detuning of the fundamental is compared
with the third harmonic in conventional FEL. The effect of optimizing
the $\beta$ function at the third harmonic in the phase shift method
is investigated in Ref. \cite{freud2} by considering the thermal
effects, where the $\beta$ function is adjusted depending on to the
energy spread. In this paper, we consider the thermal effects only
in the longitudinal direction and ignore thermal effects in the transverse
momentum, because the longitudinal spread is more effective than a
transverse spread in reducing the growth rate \cite{hafizi,freund 1991}.

The initial radiation signal in a single-pass amplifier configuration
based upon the principle of self-amplified spontaneous emission (SASE)
originates from the temporally random microbunching present on the
electron beam as they enter the wiggler. SASE amplifiers start up
from the spontaneous emission due to shot noise on the electron beam,
which is due to the random fluctuations in the phase distribution
of the electrons \cite{wang,saldin1998,Bonifacio1994}. The seed power
for the fundamental overwhelms the shot noise, whereas it treats correctly
in order to model the initial start up of harmonic radiation. Penman
and McNeil discussed a particular algorithm used in one-dimensional
(1D) approximation (in their 1D code) in order to simulate FEL start
up noise \cite{penman}. In Ref. \cite{fawley}, the shot noise algorithm
is obtained for multidimensional FEL simulation codes. Freund and
coworkers \cite{freund 2008} discussed and compared two different
shot noise models that are implemented in both 1D wiggler-averaged
(PERSEO) and non-wiggler-averaged (MEDUSA 1D) simulation codes, and
a 3D non-wiggler-averaged (MEDUSA) formulation. They found out that
there is a very good agreement between PERSEO and MEDUSA 1D for the
evolution at the fundamental and harmonics. They also showed that
the harmonics saturate slightly prior to the saturation point of the fundamental. Also, the high gain FEL, in the SASE regime, operating
with segmented undulators is studied in Ref. \cite{dottoli2} using
a set of semianalytical formulas, with the effect of the energy spread
taken into account.

The aim of this paper is to present a three-dimensional simulation
of the emission at the fundamental and third harmonic in the non-wiggler-averaged-orbit
approximation of the harmonic lasing FEL with source-dependent expansion\cite{Sprangle1,Sprangle2,Sprangle3,Sprangle4}.
Therefore, the source function is incorporated self-consistently into
the functional dependence of the radiation waist, the radiation wavefront
curvature, and the radiation amplitude instead of using the usual
modal expansion consisting of vacuum Laguerre-Gaussian or Hermite-Gaussian
functions. It is important to emphasize that no wiggler average is
imposed on the orbit equations. Therefore, It is possible to treat
the injection of the beam into the wiggler, with the ease of inclusion
of external focusing or dispersive magnetic components in the beam
line and the facility for using an actual magnetic field in the numerical
solution. The third harmonic lasing is considered so that the operating
wavelength is in the EUV domain. The slippage of the radiation with
respect to the long electron bunch is ignored. Also, we present numerical
results for the fundamental resonance and the third harmonic including
shot noise on the electron beam. We found that in many cases, SASE
FEL gives more optimistic result than seeded FEL in detuning of the
fundamental method. For example, one of the results of our studies
is that the lowest mode in shot noise is more effective at the saturation
point compared to that in the seeded FEL. 

The code for this purpose is named CYRUS 3D, which was developed by
PhD students in Amirkabir University and Institute for Research in
Fundamental Sciences (IPM)\cite{cyrus3D}. This code follows MEDUSA
3D \cite{Medusa3d} formulation.

The organization of this paper is as follows. The simulations conducted
using the CYRUS 3D simulation code is briefly described in Sec. II.
A general discussion of the retuned wiggler parameter is given in
Sec. III. The simulation results for the fundamental resonance and
the third harmonic in this system for seeded FEL and SASE FEL are
discussed in Sec. IV. In Sec. V, the summary and discussion are presented.

\section{DESCRIPTION OF THE SIMULATION CODE}

The simulation code for three-dimensional non-wiggler averaged-orbit
formulation is CYRUS 3D code, which is written in standard Fortran
95. This code is time independent with harmonics and thermal effects
taken into account. It models planar wiggler generated by magnets
with parabolically shaped pole faces for additional focusing; hence,
we can write

\[
\mathbf{B_{\textrm{w}}}(\mathbf{X})=B_{w}(z)\Biggl\{\cos(k_{w}z)\left[\hat{\mathbf{e}}_{x}\sinh(\frac{k_{w}x}{\sqrt{2}})\sinh(\frac{k_{w}y}{\sqrt{2}})+\hat{\mathbf{e}}_{y}\cosh(\frac{k_{w}x}{\sqrt{2}})\cosh(\frac{k_{w}y}{\sqrt{2}})\right]
\]

\begin{equation}
-\sqrt{2}\hat{\mathbf{e}_{z}}\sin(k_{w}z)\cosh(\frac{k_{w}x}{\sqrt{2}})\sinh(\frac{k_{w}y}{\sqrt{2}})\Biggl\},
\end{equation}
where $B_{w}$ represents the wiggler amplitude, $k_{w}=\nicefrac{2\pi}{\lambda_{w}}$
is the wiggler wave number corresponding to the wiggler period $\lambda_{w}$.
This type of wiggler has enough focusing on an electron beam in both
directions in the plane transverse. And the electromagnetic field
is represented as a superposition of Gauss-Hermit modes in the slowly
varying amplitude approximation. The vector potential of the linearly
polarized radiation field is

\begin{equation}
\mathbf{A}_{r}(\mathbf{X},t)=\underset{{\scriptscriptstyle {\scriptstyle {\scriptscriptstyle l,n,h}}}}{{\displaystyle \sum\frac{1}{2}}}e_{l,n,h}(x,y)e^{i\alpha_{h}\unit{\unit[r^{2}/]{w_{h}^{2}}}}\delta A_{l,n,h}e^{ih(kz-\omega t)}\hat{\mathbf{e}_{x}}+c.c.,
\end{equation}
where the summation indices $l$ and $n$ denote the transverse mode
structure, $h$ is the harmonic number and $e_{l,n,h}(x,y)=e^{\nicefrac{-r^{2}}{w_{h}^{2}}}H_{l}(\sqrt{2}x/w_{h})H_{n}(\sqrt{2}y/w_{h})$
is the transverse structure of each mode. Here, $H_{l}$ is the Hermite
polynomial of order $l$, $\omega_{h}$ is the spot size of the $h$th
harmonic, $\alpha_{h}$ is related to the curvature of the phase front,
$\omega$ is the fundamental frequency, and c.c. denotes the complex
conjugate. Electron trajectories are integrated using the three-dimensional
(3D) Lorentz force equations in the magnetostatic and electromagnetic
fields. It is important to emphasize that no average is performed
over the Lorentz force equation. So, Cyrus can simulate the injection
of the electrons into each segment of the wiggler.

This code like MEDUSA 3D employs nonaverage equations. The details
of the formulation is explained in Ref. \cite{Medusa3d}. We simulate
the detuning of the fundamental FEL in which the wiggler consists
of two segments. In the harmonic lasing FEL the wiggler segments have
two different magnetic field strengths but the same wavelength $\lambda_{w}$. 

The thermal effect of the electron beam on the harmonic gain is particularly
important. The kinetic theory has shown that the thermal FEL lowers
the growth rate \cite{ibanez,hafizi}. Higher harmonics are more sensitive
to the energy spread than the fundamental one \cite{Huang-1,Mcneil2,saldin2}.
It was concluded in Ref. \cite{Mcneil2} that harmonic lasing with
phase shifting is more sensitive to the emittance and the energy spread
than the harmonic lasing with detuning of the fundamental. In Refs.
\cite{freund 1991,chakhmachi}, a spread in the traverse momentum
with constant total energy is considered. They showed that a longitudinal
spread is more effective than a transverse spread in reducing the
growth rate. 

To consider effects of the energy spread, we assume longitudinal spread
without any spread in the transverse momentum. So, the initial conditions
is chosen to model the axial injection of the electron beam with the
energy in the form of a Gaussian distribution function that is peaked
around the initial energy of the beam. We choose the thermal distribution
function as

\begin{eqnarray}
G_{0}(p_{z}) & = & \sqrt{\frac{2}{\pi}}\frac{1}{\Delta p_{z}}exp\left(-\frac{2\left(p_{z}-p_{z0}\right)^{2}}{\Delta p_{z^{2}}}\right),
\end{eqnarray}
where $p_{0}$ and $\Delta p_{z0}$ are the initial bulk momentum
and momentum spread, respectively. The corresponding axial energy
spread can be written as

\begin{eqnarray}
\frac{\Delta\gamma_{0}}{\gamma_{0}} & = & \frac{\gamma_{0}^{2}-1}{\gamma_{0}^{2}}\frac{\Delta p_{z0}}{p_{0}},
\end{eqnarray}
where $\gamma=\sqrt{1+p_{0}^{2}/m_{e}^{2}c^{4}}$ is the average initial
energy of electrons at the injection time. Therefore, the averaging
operator in Eq. 10 of Ref. \cite{Medusa3d} is defined by

\begin{equation}
\left\langle \left(\cdots\right)\right\rangle =\intop\frac{d\psi_{0}}{2\pi}\sigma_{\parallel}(\psi_{0})\iintop dx_{0}dy_{0}\sigma_{\perp}(x_{0},y_{0})\intop dpG_{0}(p_{z})(\cdots).
\end{equation}

Shot noise is a random component in the initial phase $\psi_{0}$
of the macroparticles in such a way that $\left|\left\langle exp(i\psi_{0})\right\rangle \right|=1/\sqrt{N_{e}}$,
where $N_{e}$ is the number of correlated electrons. In the steady-state
simulation, each beamlet interacts with the electromagnetic field
in an identical manner so that the simulation treats only one beamlet.
Therefore, the number of electrons per beamlet $n_{e}$ is given by

\begin{equation}
n_{e}=\frac{I_{b}\lambda}{ev_{b}},
\end{equation}
where $I_{b}$ is the beam current, $\lambda$ is the wavelength,
$e$ is the electron charge, and $v_{b}$ is the bulk axial beam velocity.
Since the total number of interacting electrons $N_{e}$ will include
contribution from multiple slices it will be given approximately by
\cite{Giannessi2}

\begin{equation}
N_{e}=4.3\frac{L_{g}}{\lambda_{w}}n_{e},
\end{equation}
where $L_{g}$ is the field exponentiation length.

In order to introduce shot noise, a perturbation with the desired
harmonic content is added to the initial phase $\left\{ \psi_{0j}\right\} $
to obtain a distribution of $\left\{ \psi_{0j}^{\prime}\right\} $,
which can be written as \cite{freund 2008}

\begin{equation}
\psi_{0j}^{\prime}=\psi_{0j}+\underset{h}{\sum}\delta\psi_{h}\sin\left[h(\psi_{0j}-\varphi)\right],
\end{equation}
where $\delta\psi_{h}\ll1$ is chosen to describe the Poisson statistics
and $\varphi$ is chosen randomly over the interval $\left[0,2\pi\right]$.
By Choosing $h\delta\psi_{h}=\delta\psi_{1}$ and $\delta\psi_{1}=2/\sqrt{N_{e}}$
we will have

\begin{equation}
\left|\left\langle exp(ih\psi_{0}^{\prime})\right\rangle \right|=\frac{1}{2}\delta\psi_{1}=\frac{1}{\sqrt{N_{e}}},
\end{equation}
and the correct Poisson statistics is recovered.

\section{RETUNED WIGGLER PARAMETER}

We consider the retuned fundamental resonant wavelength by changes
in the wiggler magnetic field alone. The wiggler is composed of two
segments and the wavelength of the fundamental resonance of the second
segment is decreased by reducing the magnetic field strength of the
second segment of the wiggler. In this case, for the first segment,
the rms wiggler parameter is $a_{1}$ and the fundamental resonant
wavelength is $\lambda_{1}$ giving the harmonic resonant wavelengths
as $\lambda_{h}=\nicefrac{\lambda_{1}}{h}$ , $h=3,5,7,\ldots$. In
the second segment, the rms wiggler parameter is reset to $a_{n}$
so that the new resonant fundamental wavelength is the $n$th harmonic
of the first segment, $\lambda_{1}^{\prime}=\lambda_{n}$. For the
assumed fixed beam energy and wiggler period, the retuned wiggler
parameter $a_{n}$ is obtain from the FEL resonance relation
\begin{equation}
\frac{1+a_{1}^{2}}{1+a_{n}^{2}}=n.
\end{equation}

Obviously, $a_{1}$ must be larger than $a_{c}=\sqrt{n-1}$ as there
are no real solutions for $a_{n}$ for $a_{1}<a_{c}$. So, the wiggler
can not be reduced to a fundamental wavelength $\lambda_{1}^{\prime}=\lambda_{n}$
for $a_{1}<a_{c}$. We consider tuning the harmonic interaction by
decreasing the wiggler magnetic field; it is clear that this is impractical
for an operating X-ray FEL.

\section{Numerical analysis}

Self-consistent first-order nonlinear differential equations are solved
numerically using the fourth-order Runge-Kutta algorithm subject to
the appropriate initial conditions and in the time independent approximation
where the pulse length is much longer than the slippage length over
the course of the wiggler. The particle averages are carried out using
a Gaussian quadrature technique in each of the degrees of freedom
($x_{0},y_{0},\psi_{0},\varphi_{0},p_{z0},$$\gamma_{0}$). The number
of Gauss-Hermite modes that are needed in the code depends on each
particular example. The self-guiding effects of the electron beam
in an FEL during exponential gain become dominant over the diffraction,
and the balance depends on the Rayleigh length, the growth rate, and
the evolution of the beam envelope. Therefore, it is necessary to
choose a suitable basis set in order to determine the optical mode
content. The number of modes that are determined by an empirical procedure
in which successive simulation runs are made with an increasing number
of modes until convergence of the saturation power and saturation
length are achieved.

The parameters for the electron beam, the wiggler, and the radiation
in the simulation are as follows. The electron beam has the relativistic
factor of 964, a peak current of 300 $A$, an initial radius of 0.01495
$cm$, and an energy spread of 0.1\% . The wiggler period is 3.3 $cm$
and exhibits a peak of on-axis amplitude equal to 10.06 $KG$. An
entry taper region is $N_{w}=10$ wiggler periods in length which
is necessary in order to inject the electrons into the steady-state
trajectories. Using these beams and wiggler parameters, the fundamental
resonance is at a wavelength of 102.9 $nm$ in the 1D resonance formula.
Because of the betatron motion in three dimensions, fundamental resonance
is found at the wavelength of 103.6 $nm$, which is seeded with a
$10\:W$ of optical power. The third harmonic wavelength is at 34.5
$nm$ and starts from zero initial power. The initial radiation waists
are 0.037 $cm$ and the initial alpha parameters are chosen to be
zero. The initial state of electron beams is chosen to model the injection
of an axisymmetric electron beams with the flattop density profiles,
i.e., $\sigma_{\perp}=1$. For unbunched electron beam, the particles
are uniformly distributed in phase. 

In this case, we assume a seed power for the fundamental of $10\:KW$
and include the third harmonic, which starts from zero initial power.
The fundamental resonance is suppressed by reducing the wiggler magnetic
field strength at $L_{1}=10\,m$ with $a_{3}=0.97$ while the third
harmonic grows to saturation. In Ref. \cite{Salehi}, it has been
shown that the saturation power and saturation length of the third
harmonic radiation depends on the length of the first segment of the
wiggler $L_{1}$. Since the main bunching for the electron beam in
the first segment of the wiggler corresponds to the fundamental resonance
the electron beam in the beginning of the second segment of the wiggler,
in which the third harmonic of the first segment of the wiggler is
a seed for the second segment, is not uniformly distributed. This
difference in the distribution of the electrons at different lengths
of the first segment of the wiggler leads to changes in the saturation
power and saturation length in the second segment of the wiggler,
compared to the conventional wiggler. The optimum length of the first
segment of the wiggler is determined by a successive runs of the code
until an optimized saturation power and saturation length of the third
harmonic are obtained.

\begin{figure}
\includegraphics[width=8cm,height=6cm]{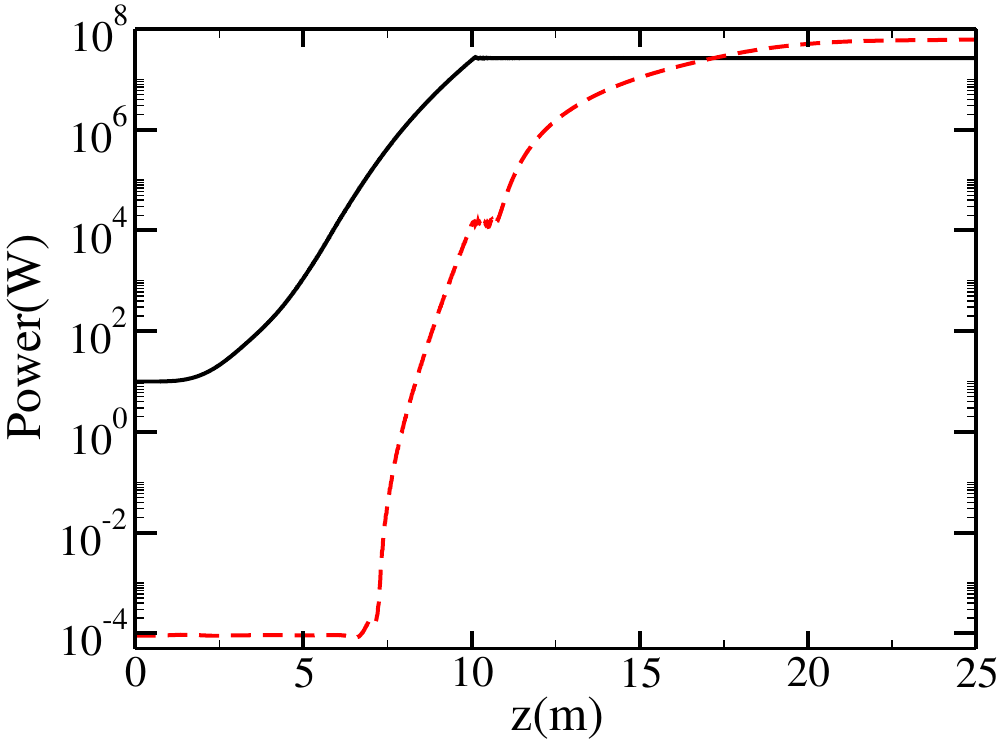}\caption{Evolution for the power for the fudamental resonance (solid line)
and the third harmonic (dashed line).\label{fig:power}}
\end{figure}

In Fig. \ref{fig:power}, the power of the fundamental resonance (solid
line) and the third harmonic (dashed line) are plotted as functions
of the distance through the wiggler. The intensity of the shorter
wavelength is larger than the intensity of the fundamental wavelength.
This means that by reducing the wiggler magnetic field, the fundamental
resonance will be suppressed and the third harmonic of the first segment
of the wiggler is a seed for the fundamental harmonic of the second
segment of the wiggler leading to higher power. The fundamental resonance
in Fig. \ref{fig:power} is suppressed at $z=10\,m$ with the power
of $2.7\times10^{7}\,W$. The third harmonic has three distinct regimes,
a small gain regime that ends at $z=6.6\,m$, exponential growth,
and approach to the saturation point at $z=21.5\,m$ with the power
of $4.8\times10^{7}\,W$.

\begin{figure}
\includegraphics[width=7cm,height=6cm]{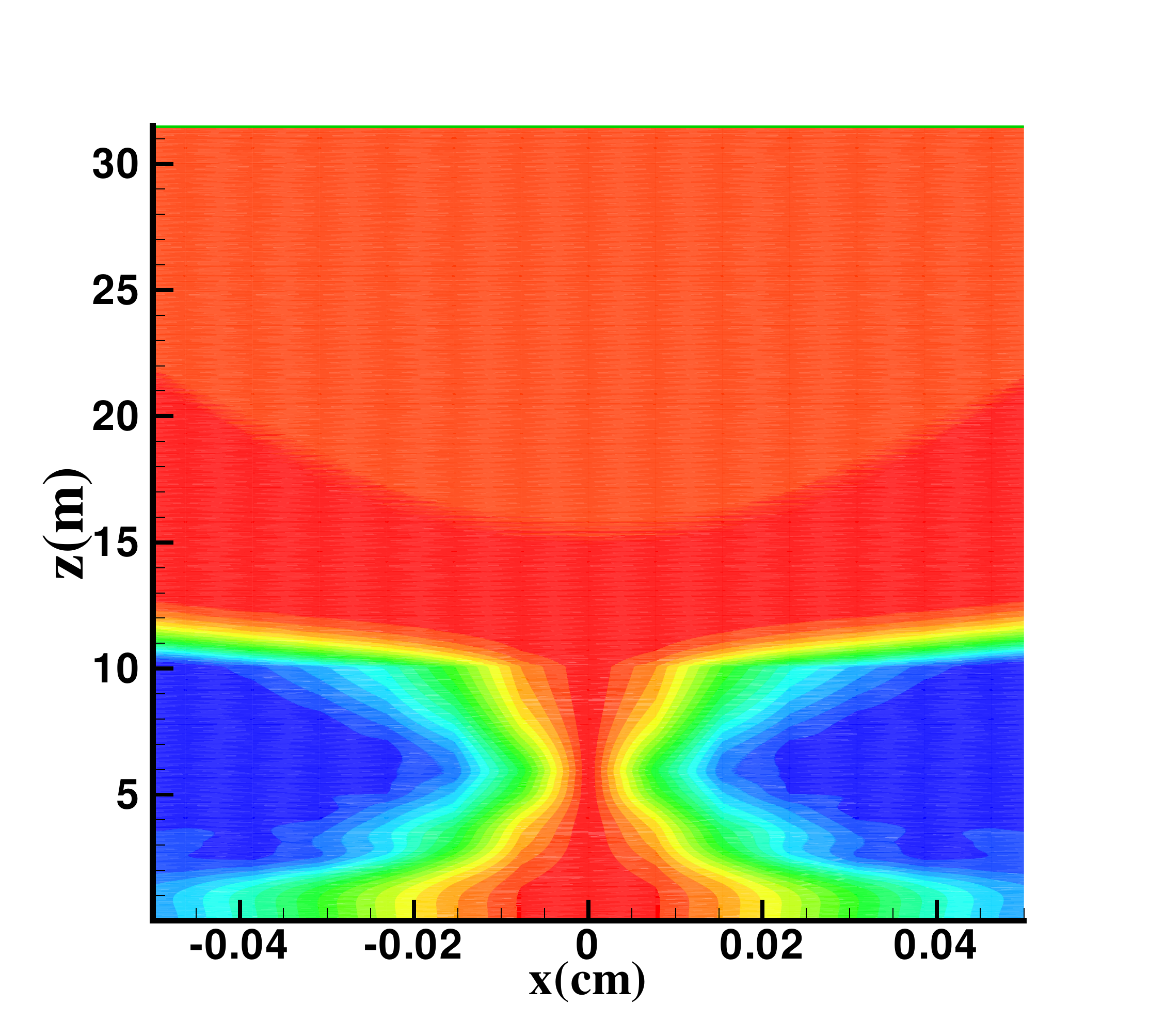}\includegraphics[width=7cm,height=6cm]{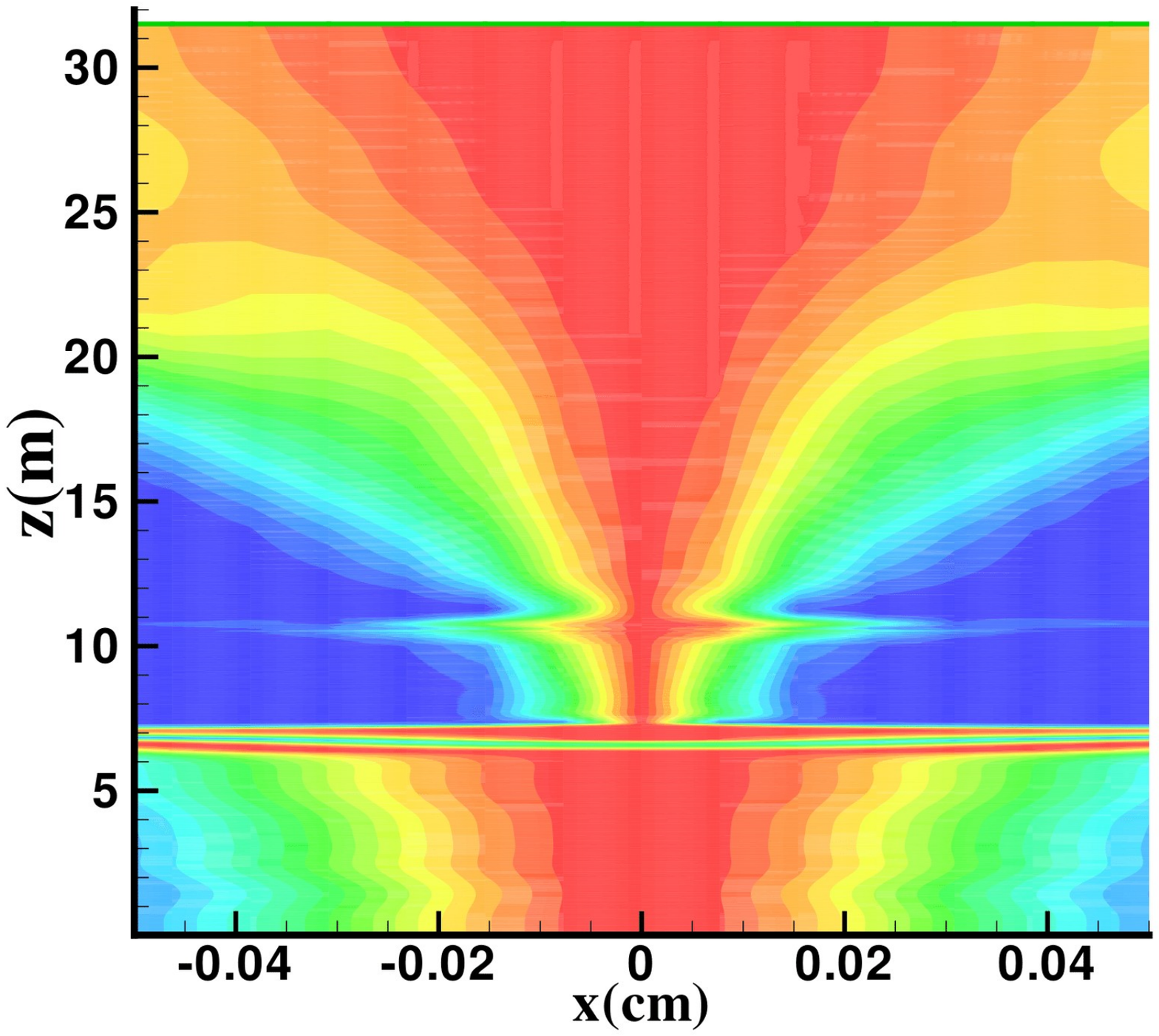}

\caption{Transverse intensity profile of the fundamental resonance wavelength
(a) and the third harmonic wavelength (b) in the x direction for $y=0$.\label{fig:Transverseprofile}}
\end{figure}

\begin{figure}
\includegraphics[width=5cm,height=4cm]{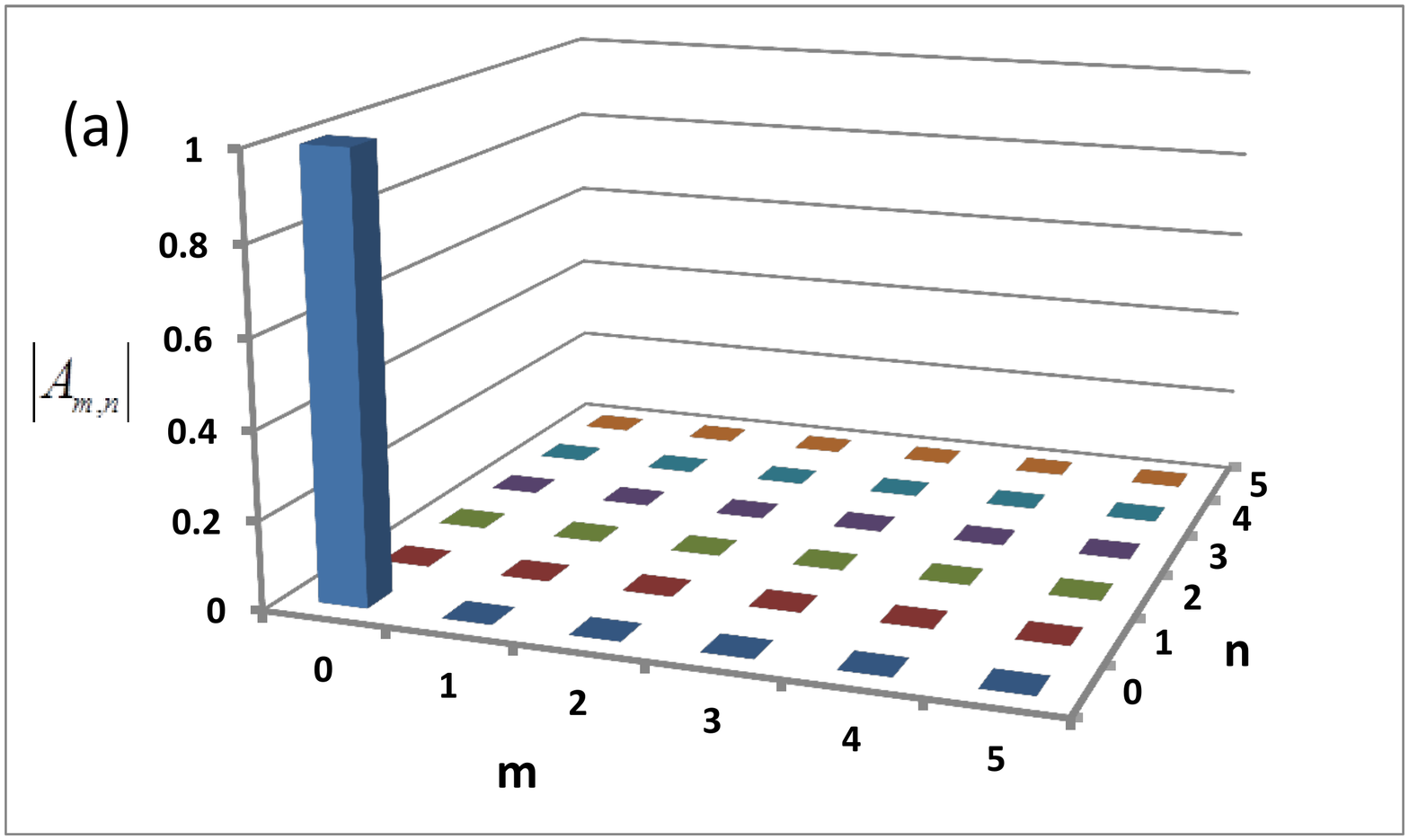}\includegraphics[width=5cm,height=4cm]{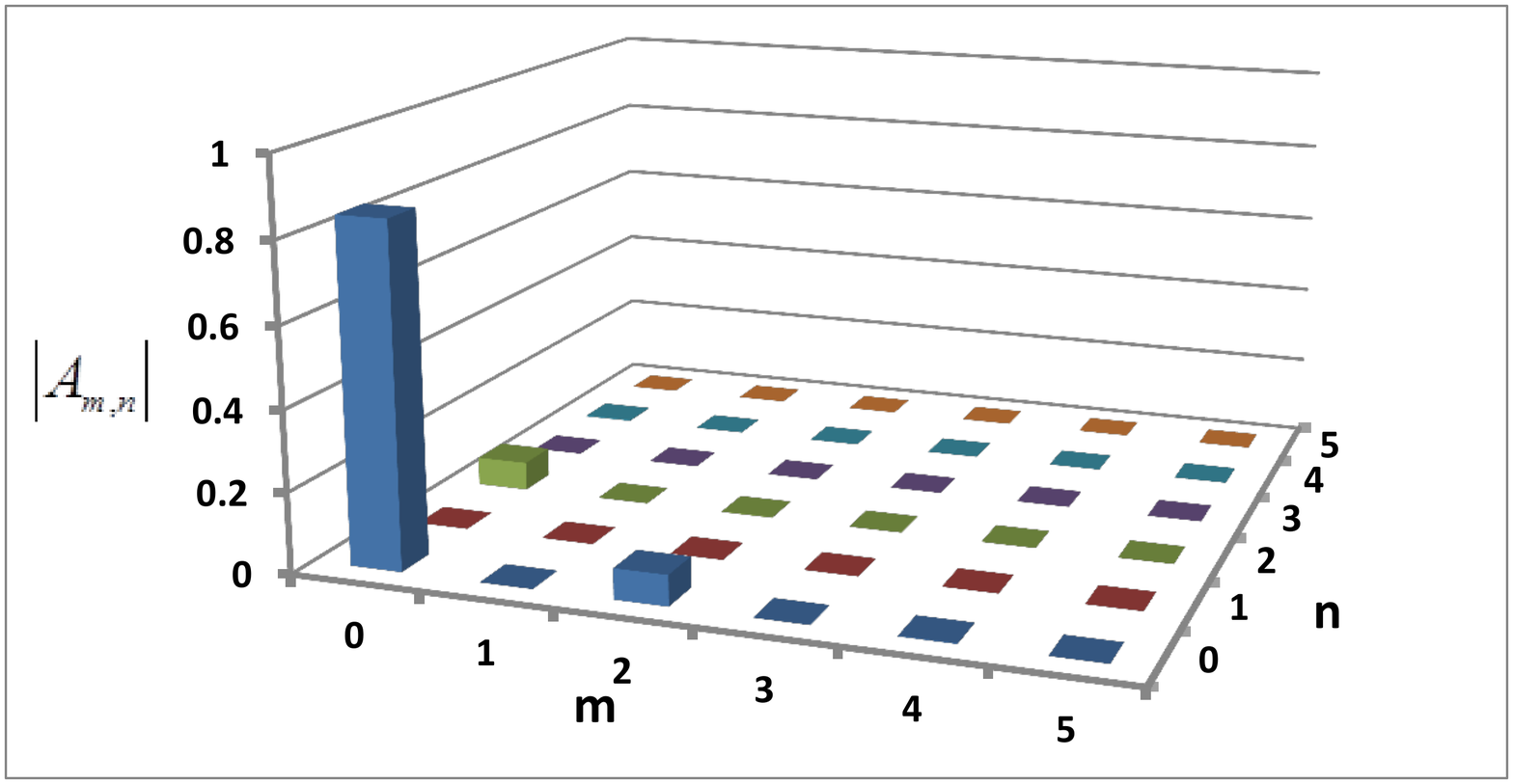}\includegraphics[width=5cm,height=4cm]{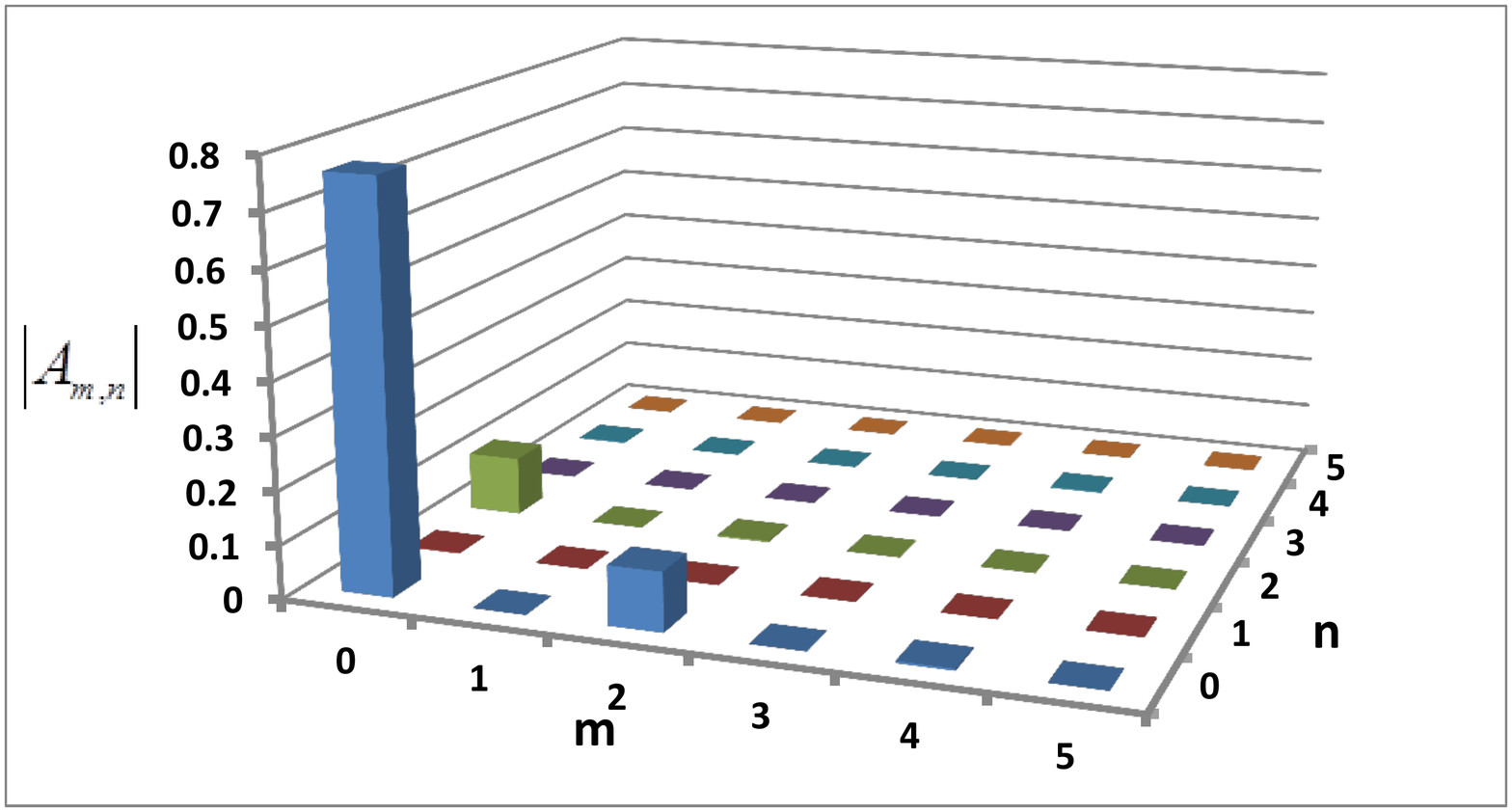}

\caption{Composition histogram of modes of the third harmonic at the beginning
of the wiggler(a), the suppresion point (b), and the saturation point
(c).\label{fig:Composition-histogram}}
\end{figure}

Evolution of the radiation amplitude in the transverse plane is shown
in Figs. \ref{fig:Transverseprofile}(a) and \ref{fig:Transverseprofile}(b)
as a function of z for the fundamental mode and the third harmonic,
respectively. Due to the normalization of transverse profile to peak
intensity of 1, these figures do not show the amplification of the
radiation; however, they show that the profile of the amplitude of
radiation in the transverse plane gets narrower when the radiation
moves toward the saturation point. It needs to be mentioned that this
mode narrowing is stronger for the third harmonic. Figure \ref{fig:Transverseprofile}(b)
shows that the transverse intensity profile of the third harmonic
initially widens up to the point where the small signal gain ends
at $z=6.6\,m$. The radiation undergoes diffraction in the small gain
region; it also experiences rapid focusing due to the gain guiding
at the onset of the exponential growth \cite{moore,scharlemann},
leading to the narrowing of the transverse intensity profile. Therefore,
it seems that the transverse profile of the radiation is guided by
the exponentially growing amplitude. In order to find the more precise
position of the saturation point, rather than using the intensity
profile of Fig. \ref{fig:power}, we can look at the point where mode
narrowing stops and the intensity profile widens. After the saturation
point the radiation waist begins to grow as the gain guiding is no
longer effective there. Before the saturation point, strong Gaussian
profiles with gentle ripples on their outskirts indicate that $TEM_{00}$
is dominant in this region. This result is also shown in Fig. \ref{fig:Composition-histogram}
in more detail. However, additional modes tend to grow, beyond the
saturation point.

\begin{figure}
\includegraphics[width=8cm,height=6cm]{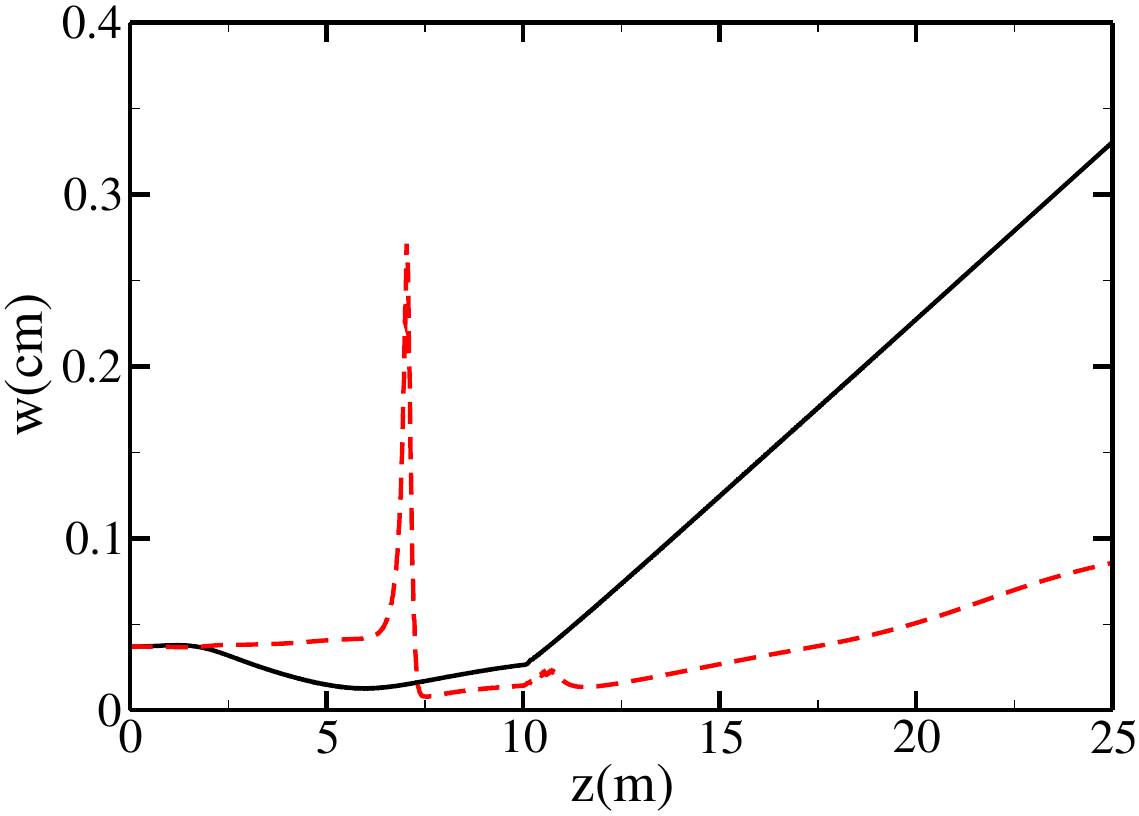}

\caption{Evolution of the radiation spot size for the fundamental resonance
(solid line) and third harmonic (dashed line).\label{fig:spotsize}}
\end{figure}

\begin{figure}
\includegraphics[width=8cm,height=6cm]{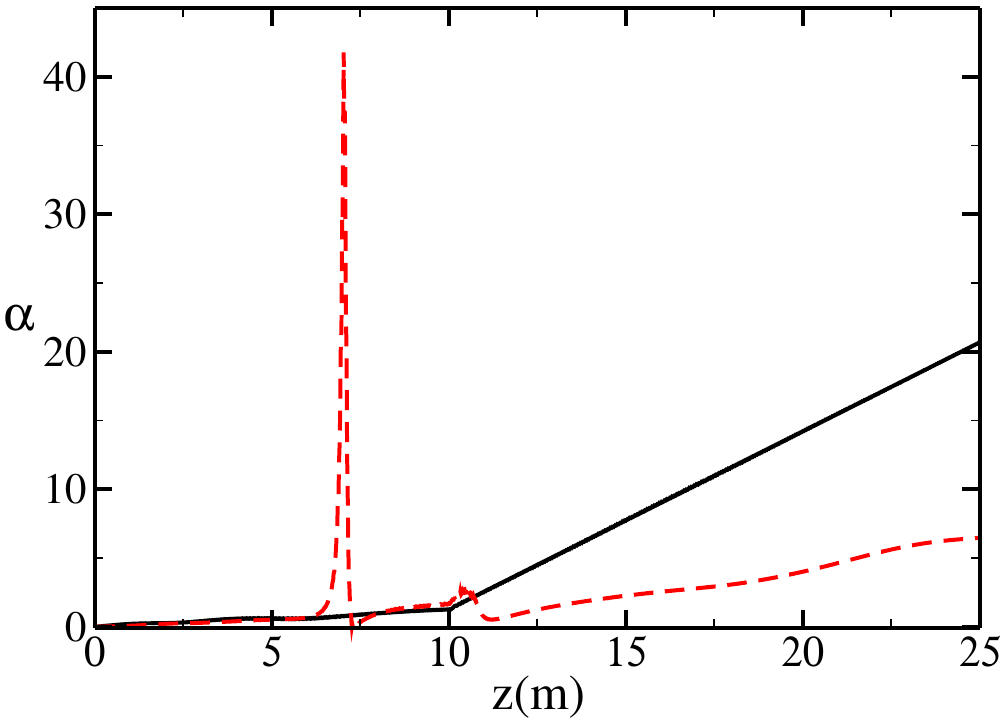}

\caption{Evolution of $\alpha_{1}$ (solid line) and $\alpha_{3}$ (dashed
line) with longitudinal coordinate.\label{fig:alpha}}
\end{figure}

In Fig. \ref{fig:spotsize}, the radiation waist of the fundamental
resonance and third harmonic are plotted. The radiation waist for
the third harmonic is observed to expand, from its initial size, in
the small signal region because of vacuum diffraction. This can also
be seen in Fig. \ref{fig:Transverseprofile}(b). In the exponential
growth region, optical guiding becomes strong and focusing is rapid.
Finally, the radiation waist expands rapidly as the saturation point
is reached. The radiation waist for the fundamental resonance behaves
to some extent differently, compared to the third harmonic, and grows
faster at the suppression point. Here, the radiation waist of the
third harmonic is smaller than that of the fundamental resonance in
the exponential growth region but the radiation waist of the third
harmonic is larger than that of the fundamental resonance in the exponential
growth region in the conventional FEL \cite{Medusa3d}.
 The curvature of the phase front,$\alpha$, is shown in Fig. \ref{fig:alpha}.
Both fundamental and the third harmonic, which are plane waves at
the entrance to the wiggler at $z=0$, deviate from plane waves as
radiation travels along the wiggler. The curvature of the phase front
of the third harmonic increases abruptly as saturation occurs but
for the fundamental resonance, it increases rapidly at the suppression
point.

It is interesting to study the composition of the optical modes (effective
higher modes ) of the third harmonic. The composition process will
introduce higher-order modes. Figure \ref{fig:Composition-histogram}
shows the mode content histogram at the beginning, suppression and
saturation point of the third harmonic. Figure \ref{fig:Composition-histogram}
indicates that a purely (0,0) mode exists at the beginning of the
wiggler.

\begin{figure}

\includegraphics[width=8cm,height=6cm]{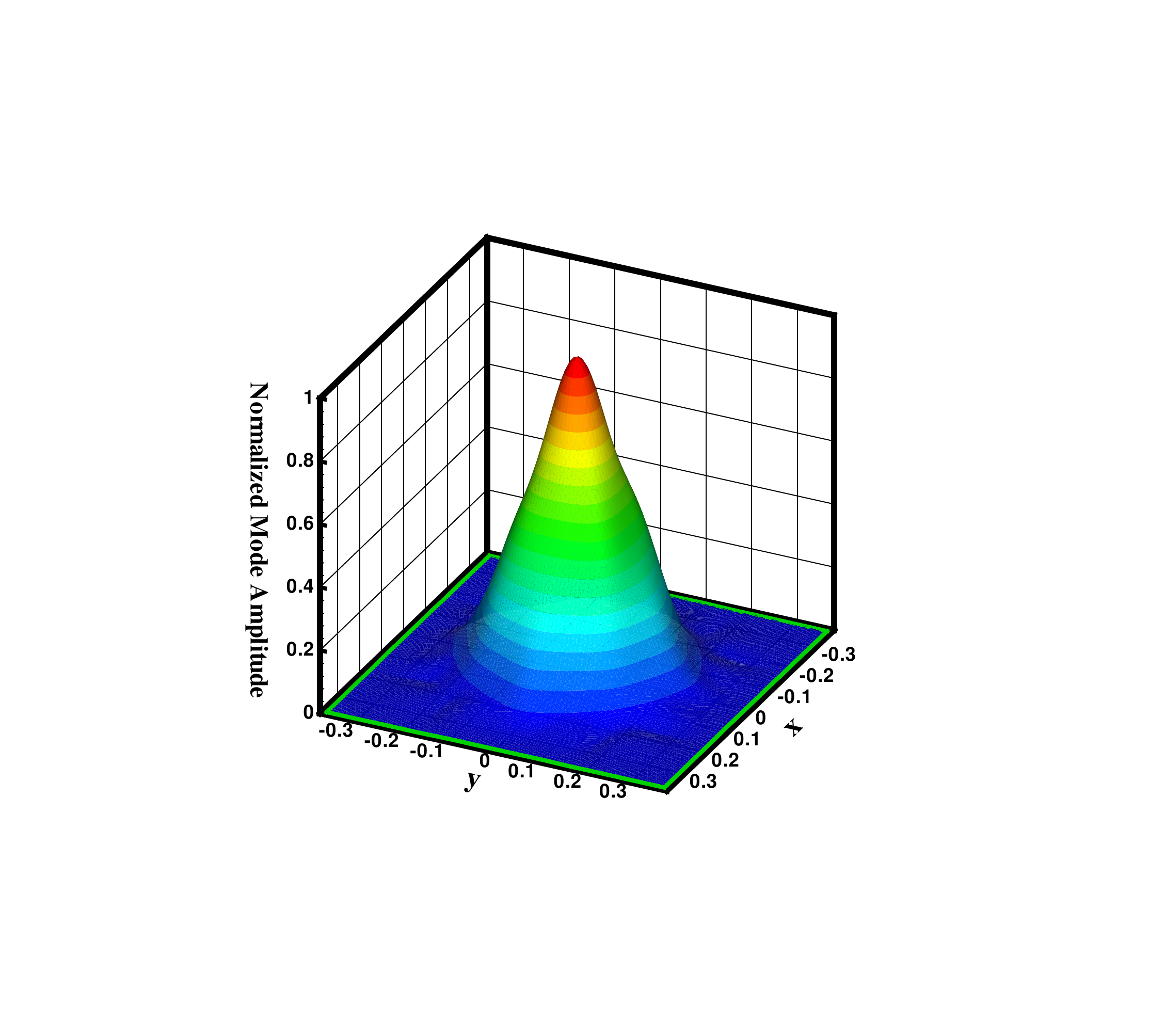}

\caption{Transverse mode pattern for the third harmonic at the saturation point
(for the third harmonic)\label{fig:amplitude}}

\end{figure}

\begin{table}
\begin{tabular}{cccc}
\hline 
 & z=0 & z=10m & z=21.5\tabularnewline
\hline 
modes & \%composition & \%composition & \%composition\tabularnewline
\hline 
(0,0) & 100\% & 84.7\% & 76.3\%\tabularnewline
(0,1) & 0.0\% & 0.0\% & 0.0\%\tabularnewline
(1,0) & 0.0\% & 0.001\% & 0.15\%\tabularnewline
(0,2) & 0.0\% & 7.6\% & 11.1\%\tabularnewline
(2,0) & 0.0\% & 7.1\% & 11.1\%\tabularnewline
(2,2) & 0.0\% & 0.11\% & 0.24\%\tabularnewline
(0,4) & 0.0\% & 0.17\% & 0.36\%\tabularnewline
(4,0) & 0.0\% & 0.18\% & 0.37\%\tabularnewline
(2,4) & 0.0\% & 0.004\% & 0.049\%\tabularnewline
(4,2) & 0.0\% & 0.004\% & 0.049\%\tabularnewline
(4,4) & 0.0\% & 0.009\% & 0.16\%\tabularnewline
\end{tabular}

\caption{Modal composition of the third harmonic in different points of the
wiggler\label{tab:Modal-composition}}
\end{table}
 Table \ref{tab:Modal-composition} shows the composition of significant
modes of the third harmonic at different points of the wiggler. It
can be seen that the composition of the lowest order mode $TEM_{0,0}$
is 84.7\% at the suppression point and decreases to 76\% at the saturation
point. It is found that the composition is 7\% for either modes $TEM_{20}$
or $TEM_{02}$ at the suppression point and increases to 11\% at the
saturation point. Therefore, the lowest order $TEM_{0,0}$ mode is
dominant and modes which are shown in Table \ref{tab:Modal-composition}
have a noticeable contribution and the rest are negligible. This higher
mode content is clearly illustrated in Fig. \ref{fig:amplitude} where
we plot the normalized intensity of the third harmonic near the saturation
point.

\begin{figure}
\includegraphics[width=5cm,height=4cm]{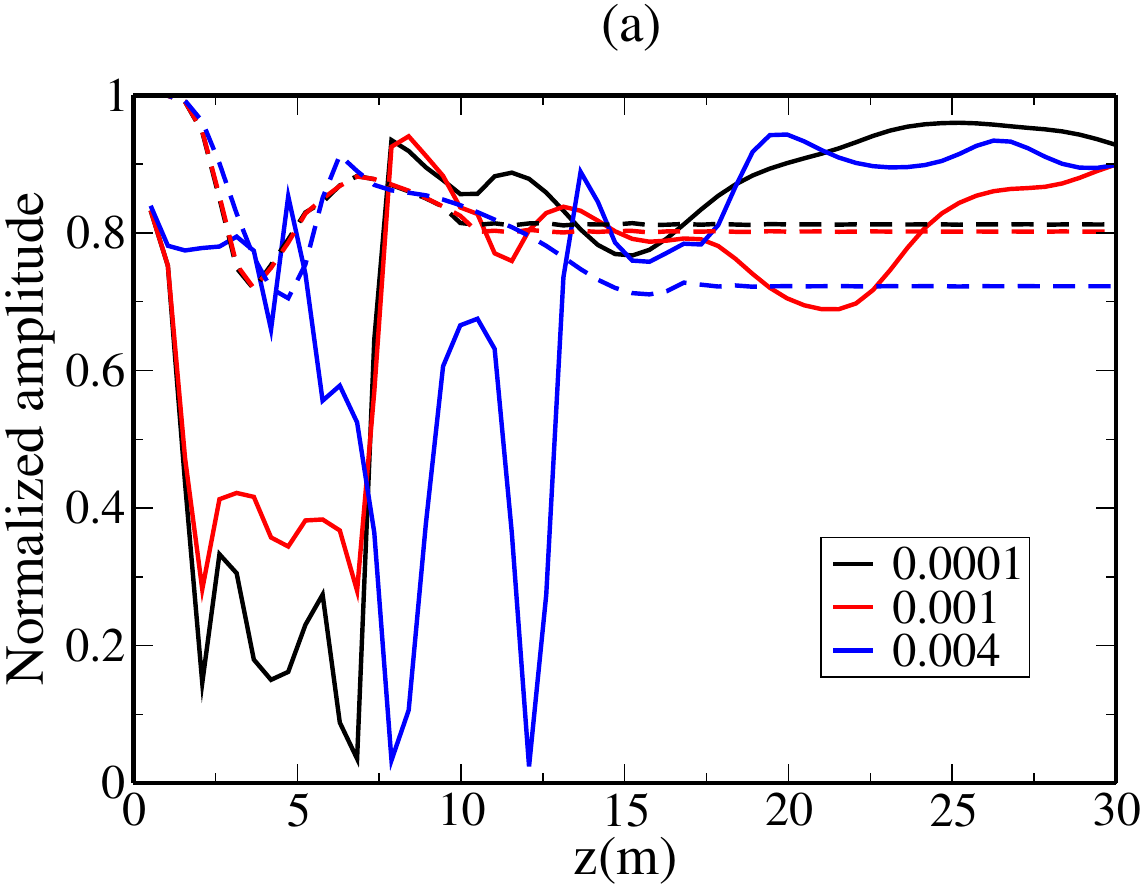}\includegraphics[width=5cm,height=4cm]{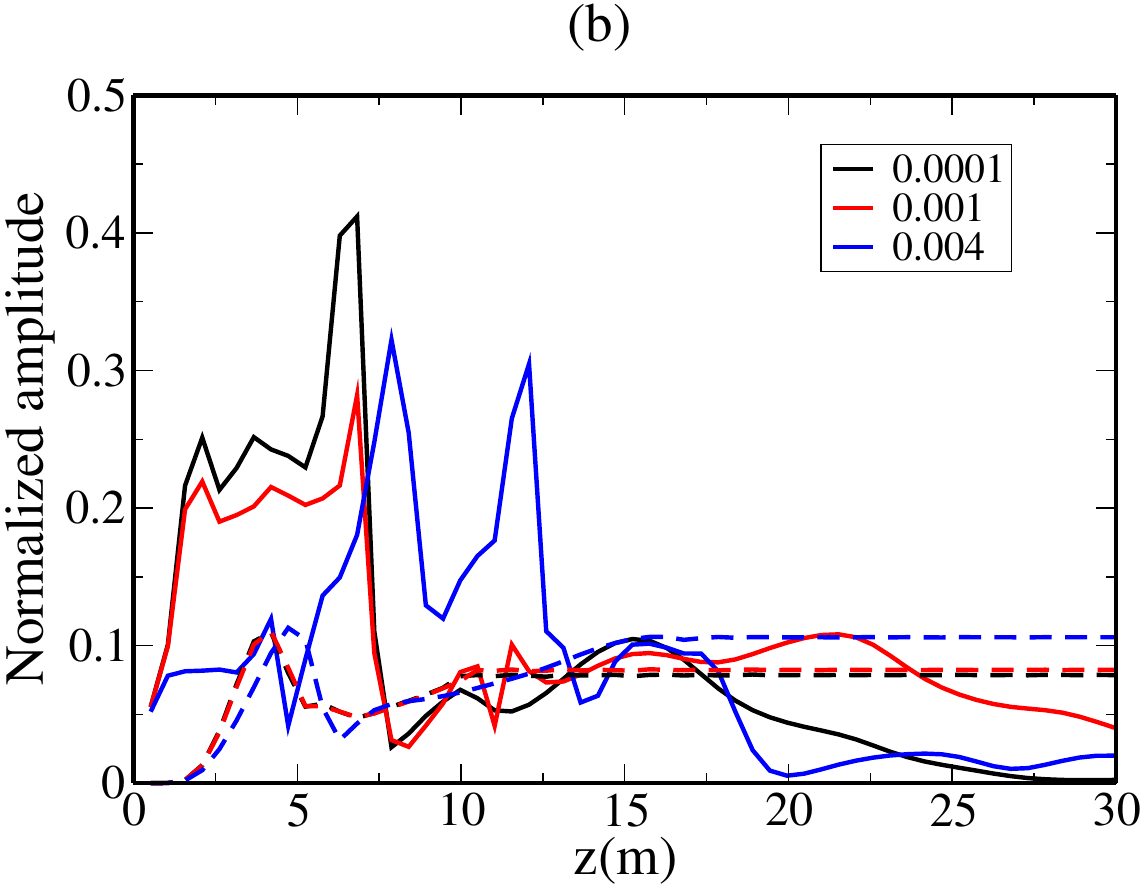}\includegraphics[width=5cm,height=4cm]{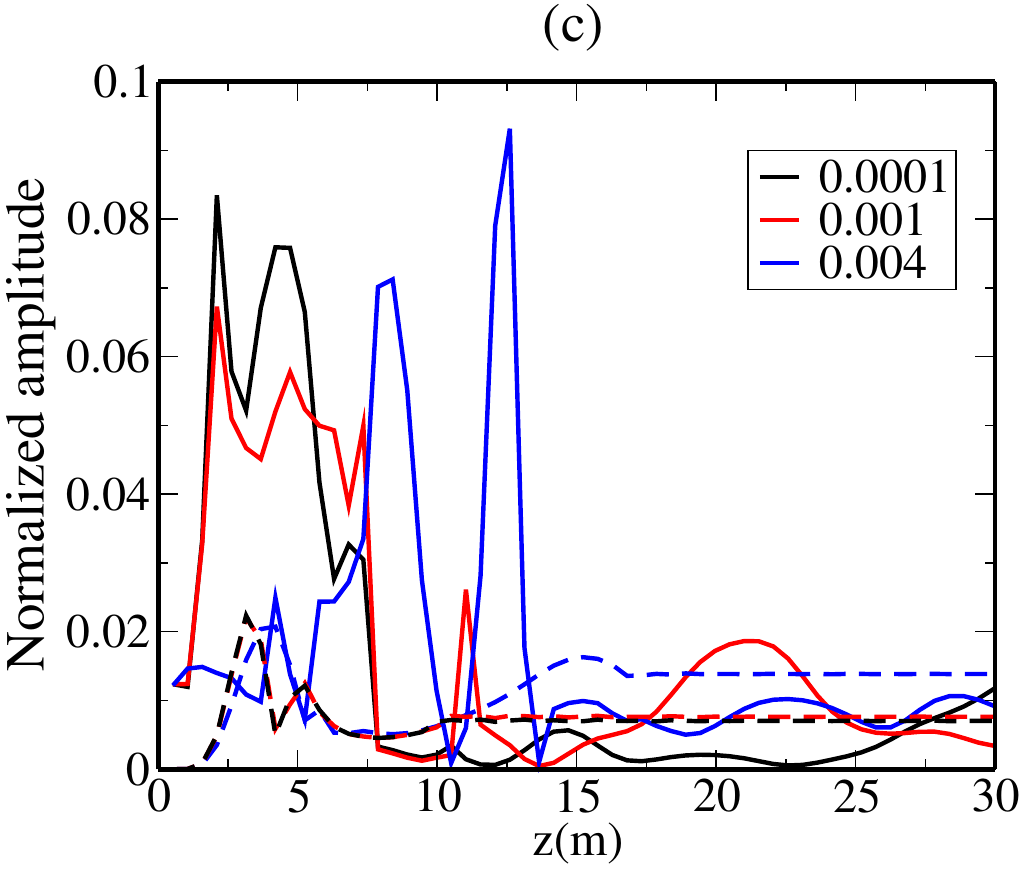}

\includegraphics[width=5cm,height=4cm]{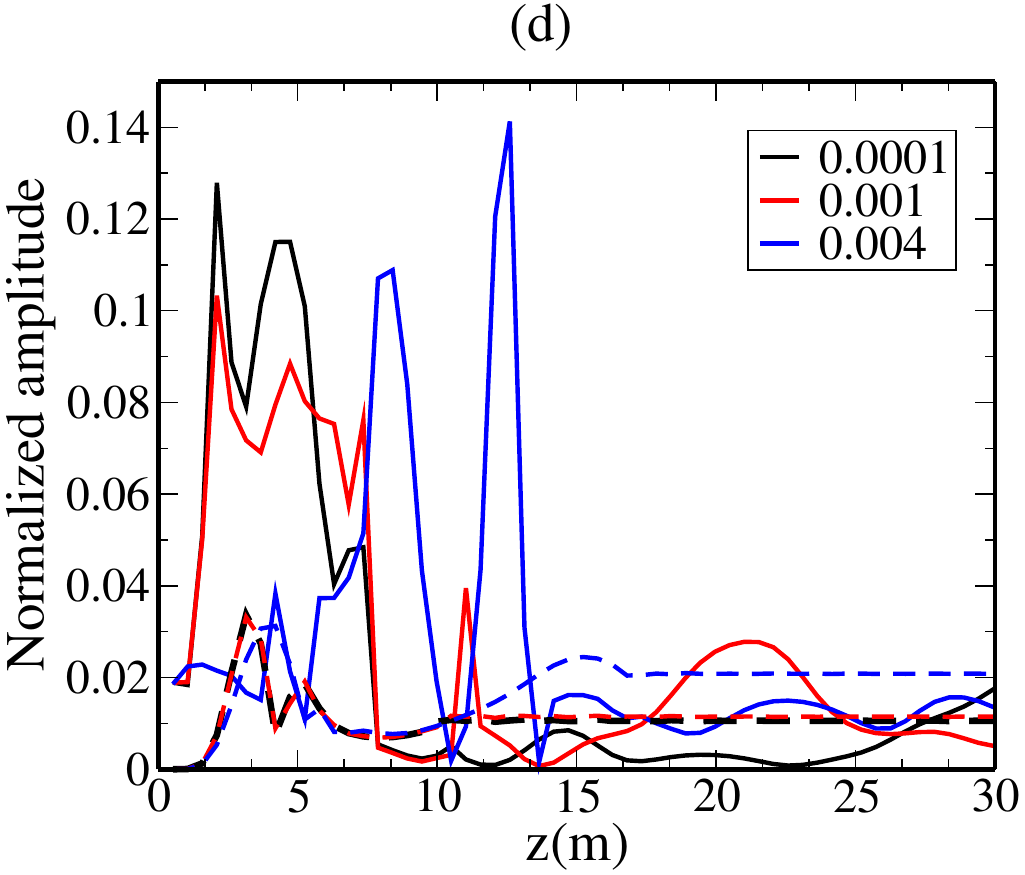}\includegraphics[width=5cm,height=4cm]{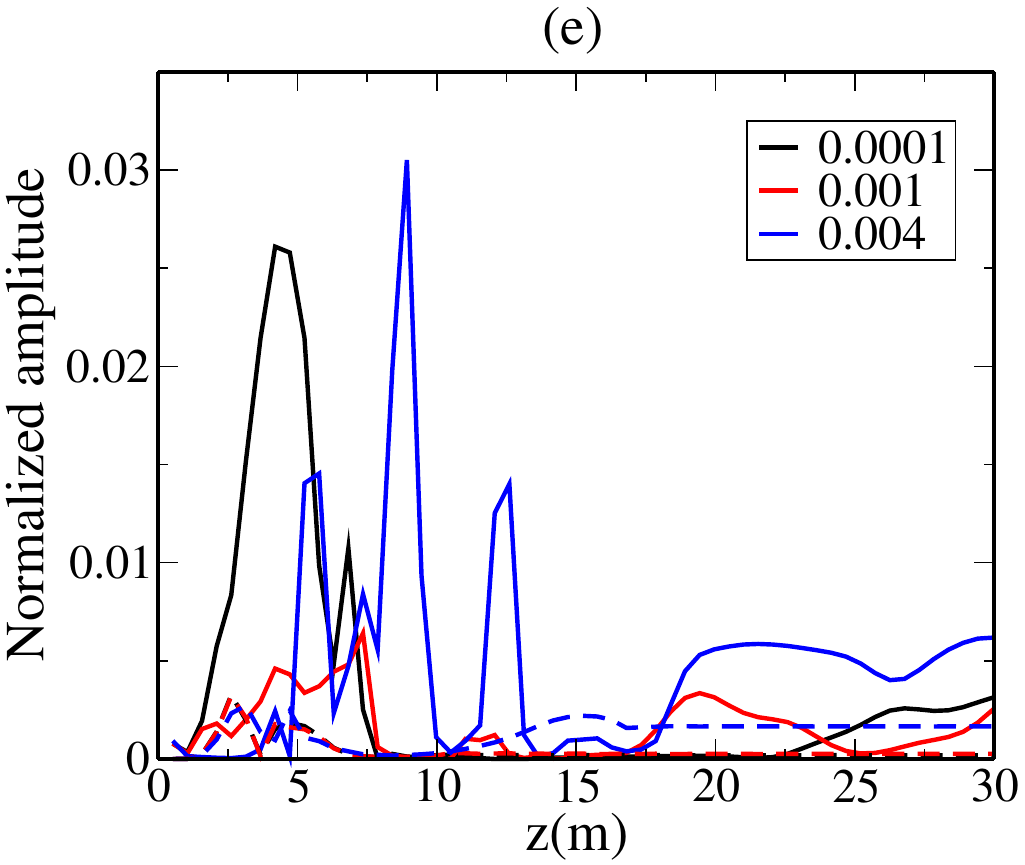}

\caption{The effect of energy spread on modes of the fundamental resonance
and the third harmonic for $\sigma_{\gamma}/\gamma=0.0001,\,0.001,\,and\,0.004$\label{fig:modes}.}
\end{figure}

The thermal effect of the electron beam on optical modes is investigated.
Figure \ref{fig:modes} shows the evaluation of the normalized amplitude
of modes (a) $TEM_{0,0}$, (b) $TEM_{0,2}$, $TEM_{2,0}$, (c) $TEM_{2,2}$,
(d) $TEM_{0,4}$, $TEM_{4,0}$, and (e) $TEM_{4,4}$ of the third
harmonic (solid line) and of the fundamental resonance (dashed line)
for initial Gaussian energy spreads of $\sigma_{\gamma}/\gamma=0.0001,\,0.001,\,and\,0.004$.
It is clear from Fig. \ref{fig:modes} that by increasing the energy
spread, intensity of the lowest mode decreases but the intensity of
higher modes increase at the saturation point. In other words, the
share of higher modes increases in higher energy spread compared to
that in the low energy spread. 

\begin{figure}
\includegraphics[width=8cm,height=6cm]{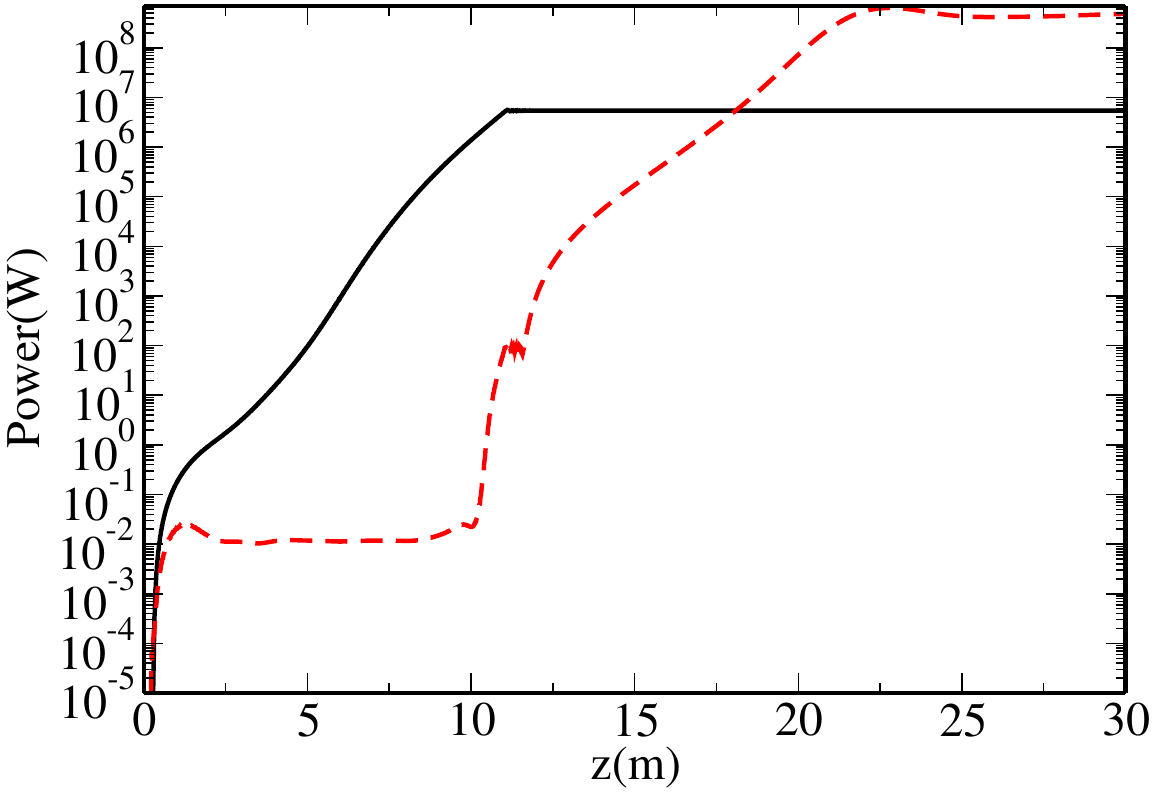}\caption{The evolution of the power in the fundamental (solid line) and the
third harmonic (dashed line).\label{powershot}}
\end{figure}

\begin{figure}
\includegraphics[width=8cm,height=6cm]{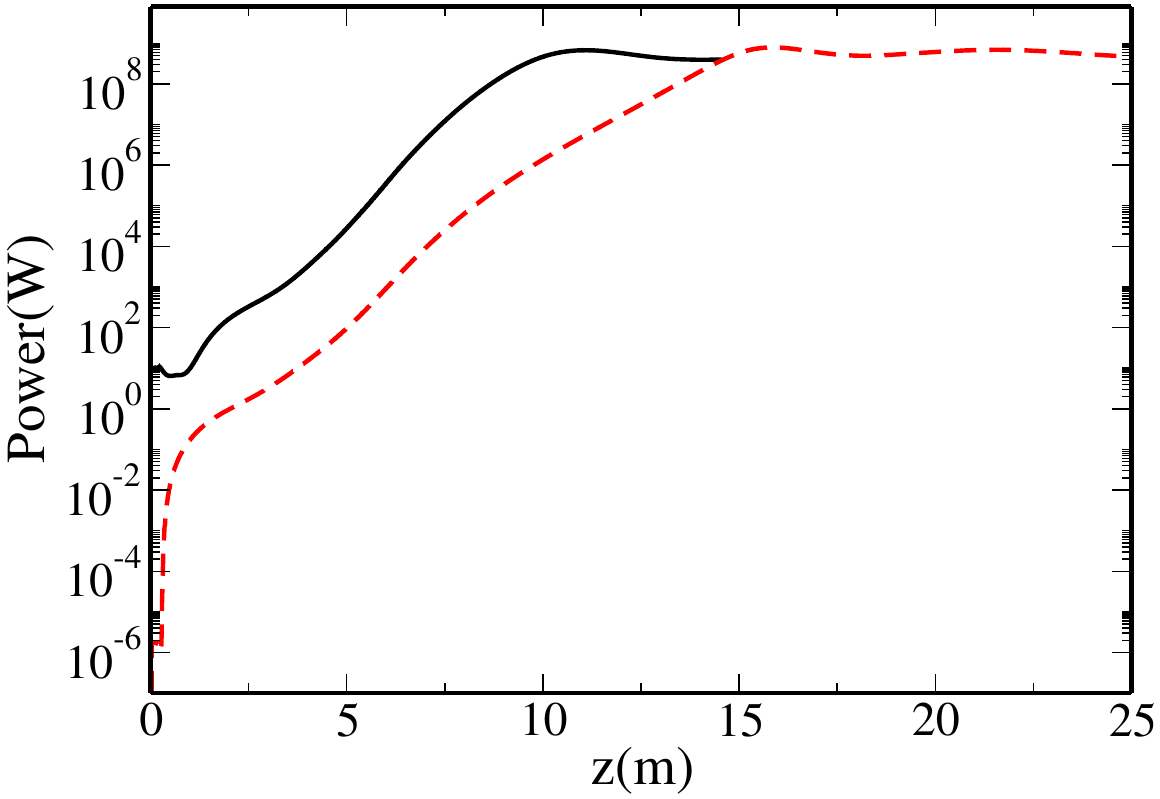}\caption{The evolution of the power of the fundamental resonance for seeded
FEL (solid line) and SASE FEL (dashed line)\label{fig:compare power}}
\end{figure}

For the shot noise, we apply an appropriate initial condition on the
initial phase of macroparticles. Also, we assume a seed power for
the fundamental of $0\,KW$ . The fundamental resonance starts from
zero and is suppressed by reducing the wiggler magnetic field strength
at $L_{1}=11\,m$. In Fig. \ref{powershot}, we plot the evolution
of powers in the fundamental and the third harmonic as a function
of the distance through the system. It can be seen that powers of
the fundamental and the third harmonic initially grow exponentially
because of start up of noise. The growth of the fundamental resonance
in Fig. \ref{powershot} is suppressed at $z=11\,m$ with the power
of $5.6\times10^{6}\,W$. The small gain regime of the third harmonic
starts from $z=1.3\,m$ to $10\,m$. 
After exponential growth regime, it saturates at $z=22.8\,m$ with
the power of $6.3\times10^{8}\,W$. We observe that the saturation
length and the saturation power of the third harmonic is considerably
increased compared to that of the case of seeded FEL. A comparison
between the seeded FEL and SASE FEL is shown in Fig. \ref{fig:compare power}
for the conventional FEL. In the conventional FEL, the saturation
length of the fundamental resonance in the seeded FEL is lower than
that of the SASE FEL.

\begin{figure}
\includegraphics[width=7cm,height=6cm]{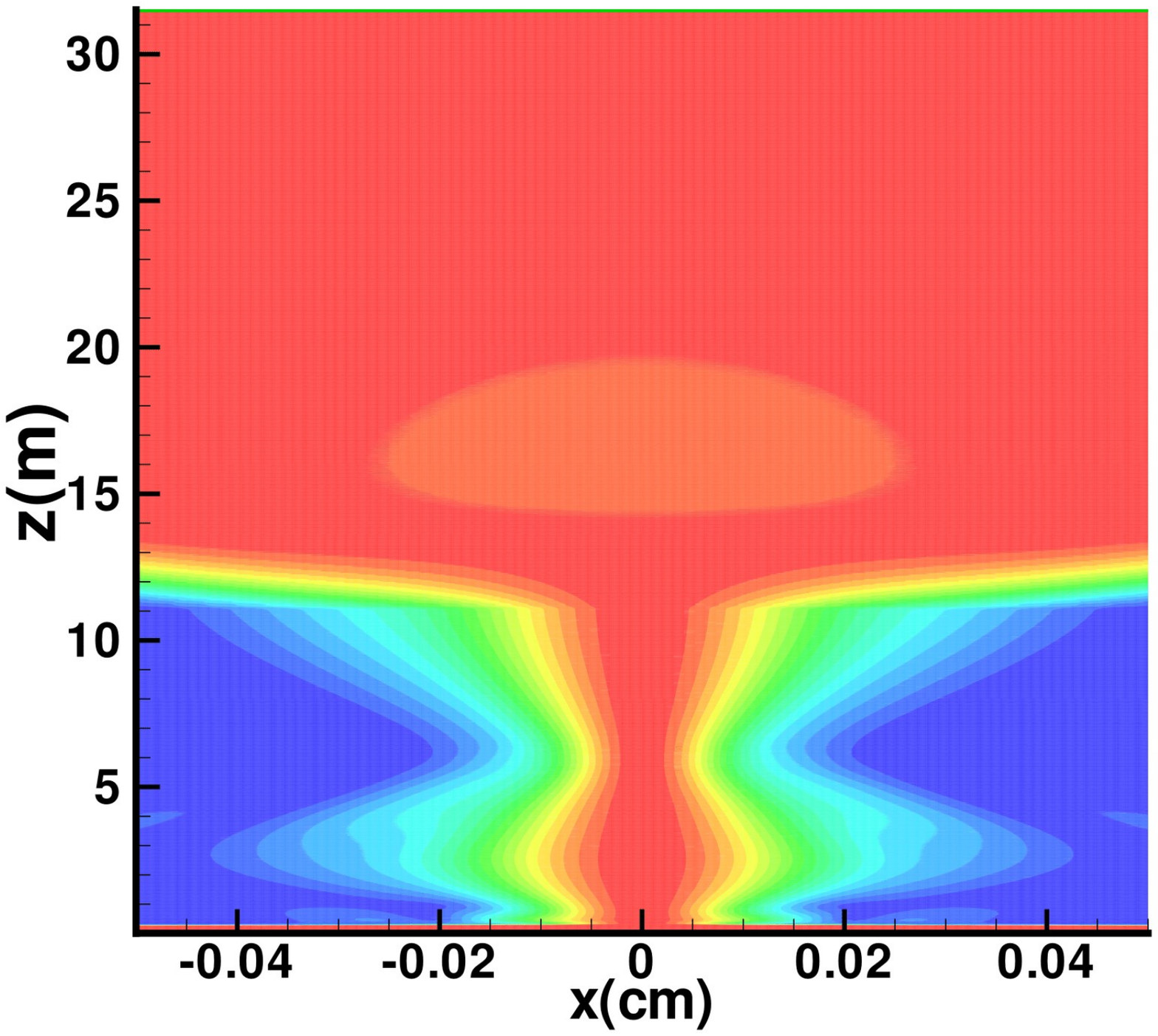}\includegraphics[width=7cm,height=6cm]{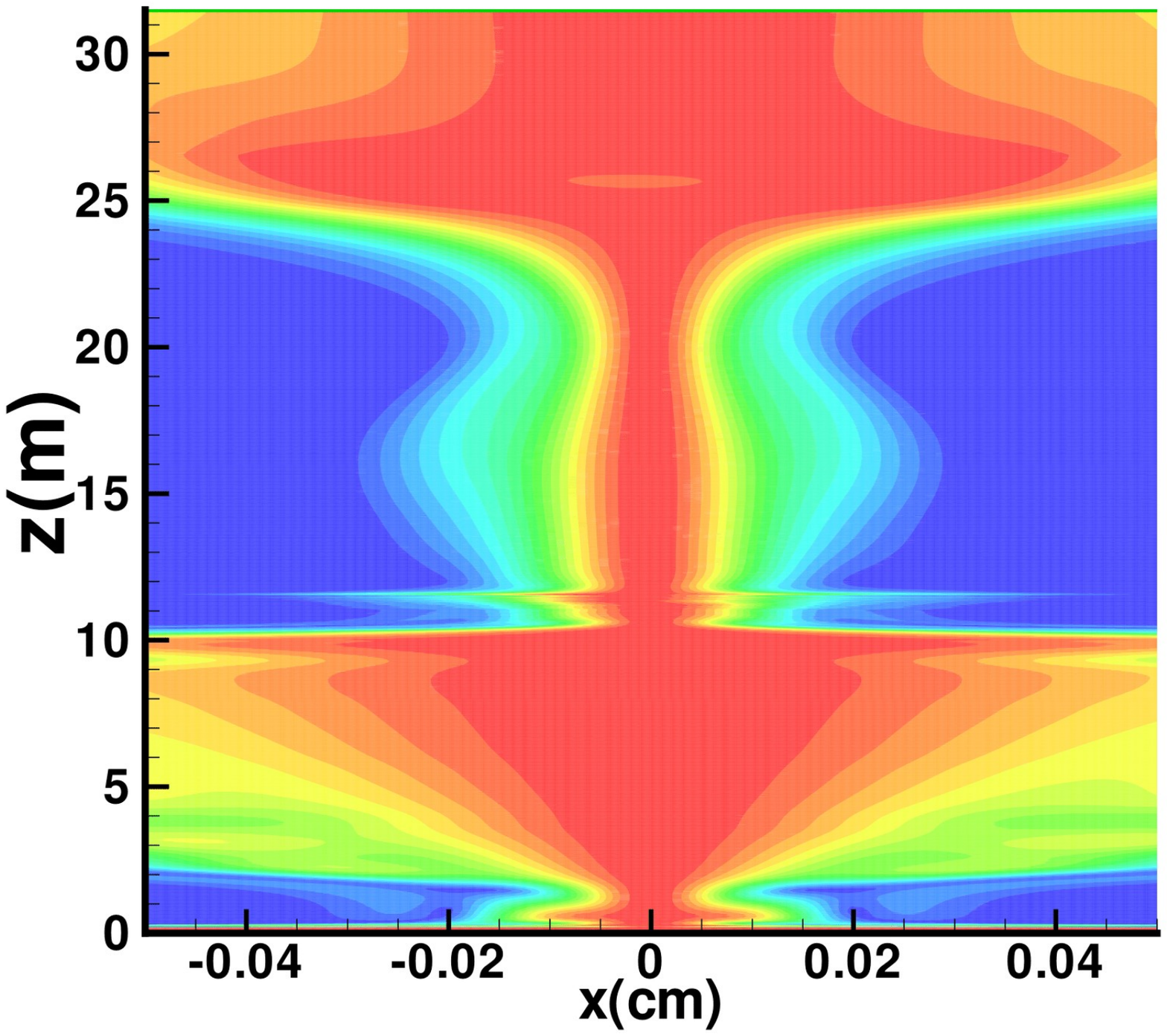}

\caption{Transverse intensity profile of the fundamental resonance wavelength
(a) and the third harmonic wavelength (b) in the y direction for $x=0$\label{fig:Transverse intensity shot}}
\end{figure}

\begin{figure}
\includegraphics[width=8cm,height=6cm]{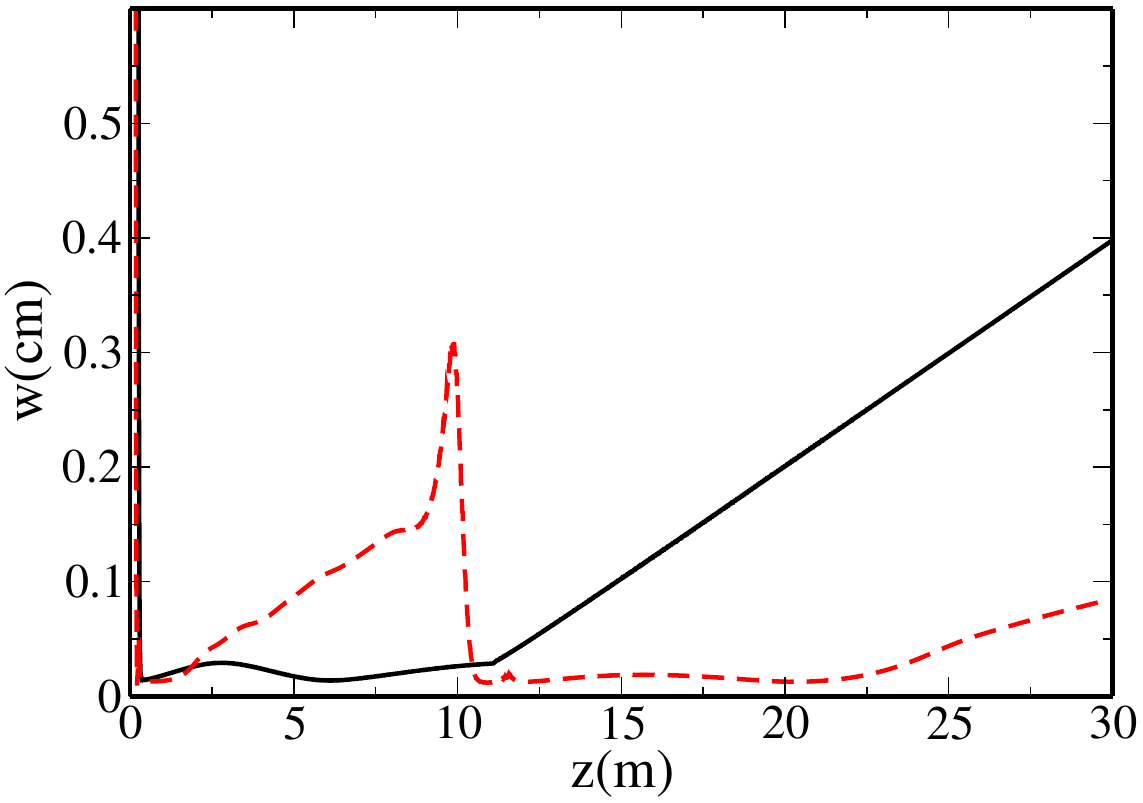}

\caption{Evolution of the radiation waist for the fundamental resonance (solid
line) and third harmonic (dashed line).\label{fig:waistshot}}
\end{figure}

\begin{figure}

\includegraphics[width=8cm,height=6cm]{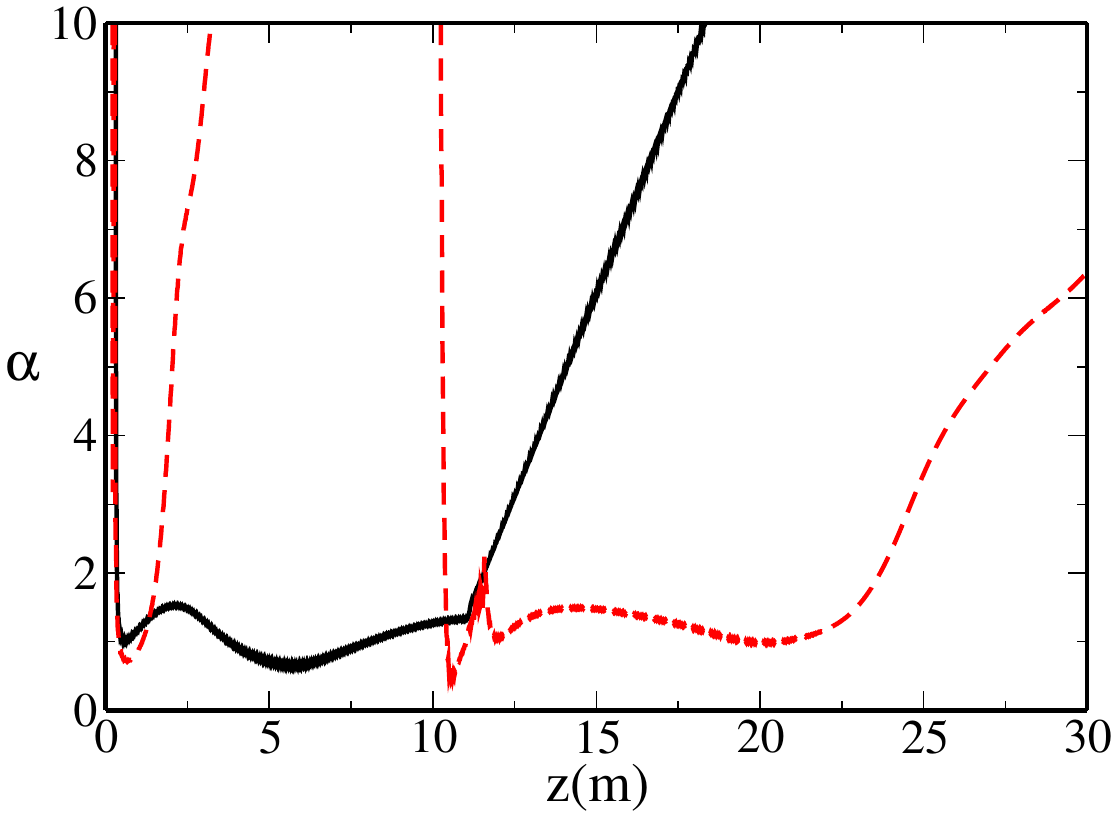}

\caption{Evolution of $\alpha_{1}$(solid line)and $\alpha_{3}$(dashed line)
with longitudinal coordinate.\label{fig:alphashot}}
\end{figure}

The transverse intensity profile and the evolution of the radiation
waist are presented in Fig. \ref{fig:Transverse intensity shot} and
\ref{fig:waistshot}.
 The effect of optical guiding due to the electron beam is evident.
As it is shown in Fig.\ref{fig:Transverse intensity shot}, the transverse
intensity is narrower than that of the seeded FEL. 
Initially, the radiation spot size for the fundamental resonance and
the third harmonic is large due to the start up from the shot noise.
In the small signal regime, the radiation waist of the third harmonic
increases. After that and at the onset of exponential growth, the
radiation waist experiences rapid focusing and decreases; it remains
small during the exponential growth before expanding at the saturation
point . The curvature of the phase front for the third harmonic at
the small signal, in Fig. \ref{fig:alphashot}, increases abruptly
but it decreases at the beginning of the exponential growth region
and behaves more like a plane wave as it moves on. Also, at the saturation
point, the curvature of the phase front is lower ($\alpha_{3}\thickapprox1.5$)
than that in the seeded FEL ($\alpha_{3}\thickapprox7$).

\begin{figure}
\includegraphics[width=5cm,height=4cm]{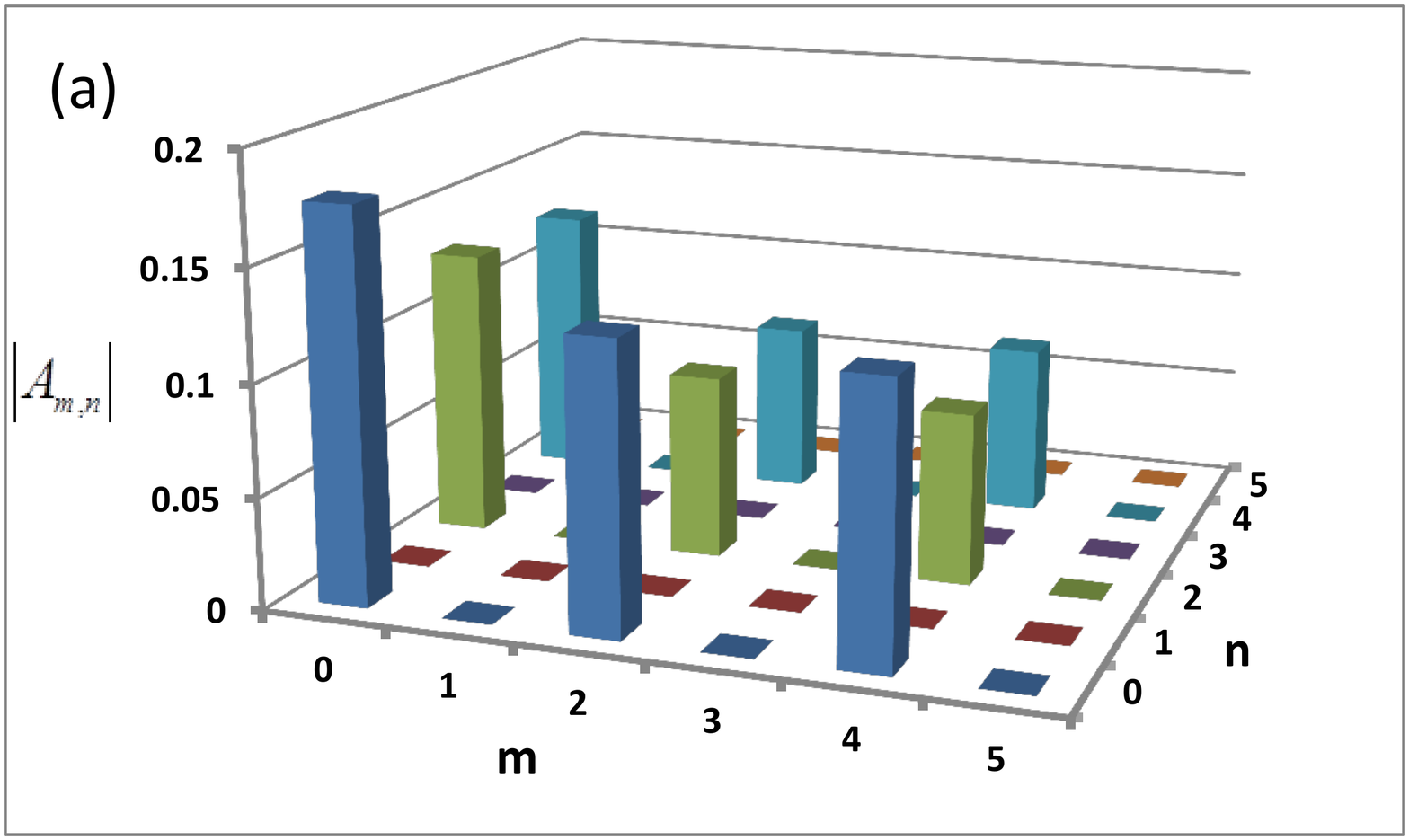}\includegraphics[width=5cm,height=4cm]{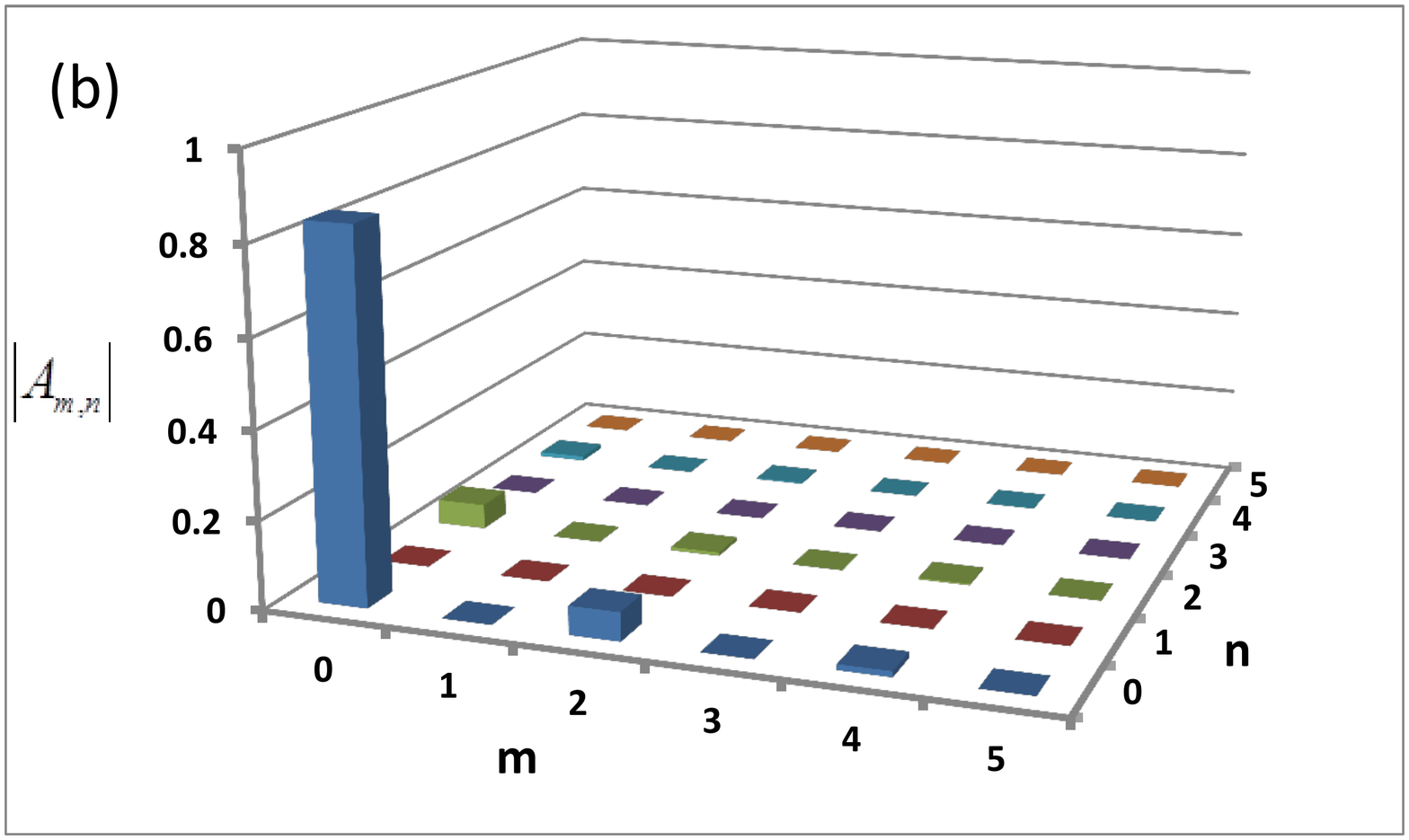}\includegraphics[width=5cm,height=4cm]{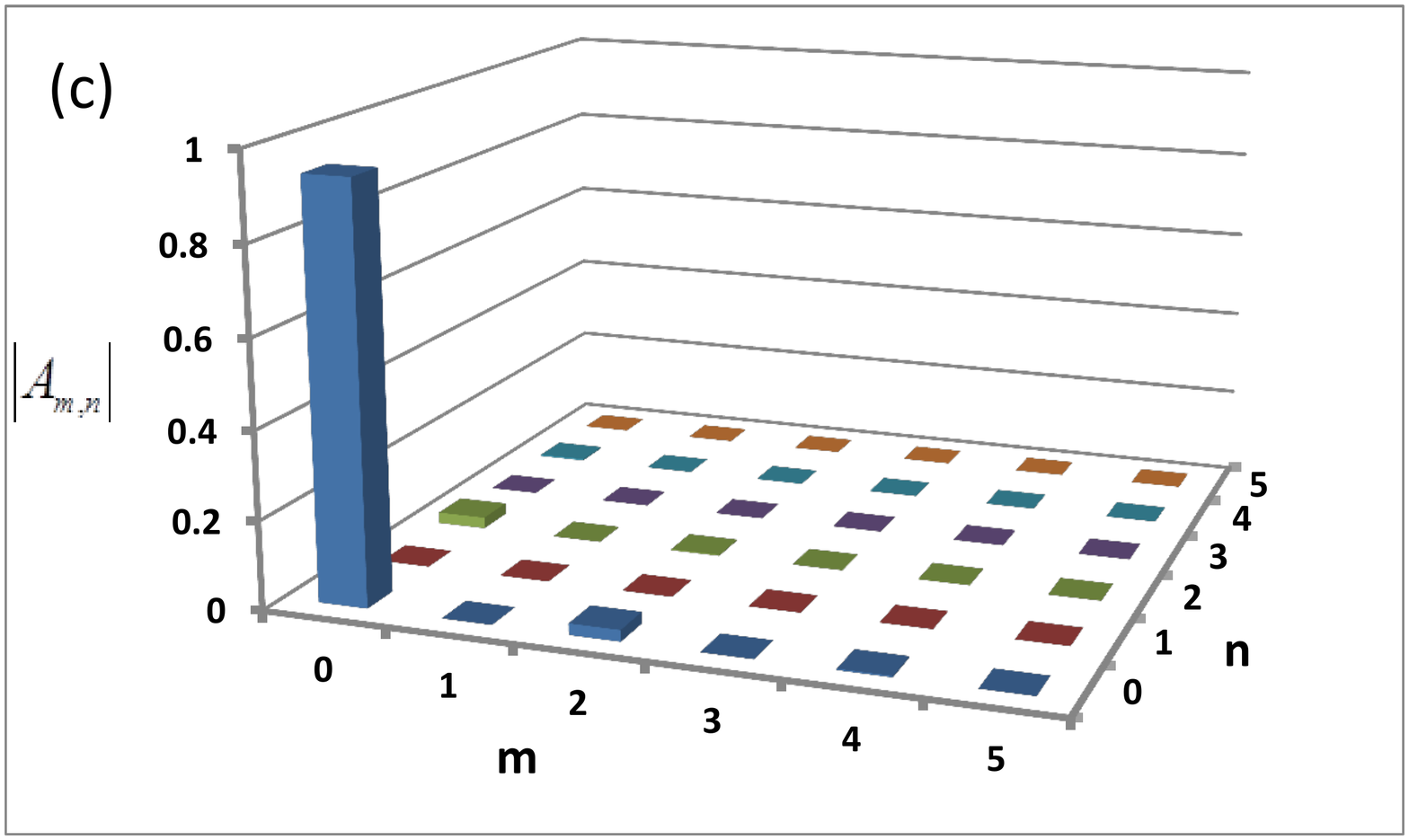}\caption{Composition histogram of modes at the beginning of the wiggler(a),
the suppresion point (b), and the saturation point (c)\label{fig:histogramshot}}
\end{figure}

The composition of the optical modes of the third harmonic at the
beginning, suppression, and saturation point is shown in Fig. \ref{fig:histogramshot}
and Tabel \ref{tab:Modal-composition-shot}. It is evident that the
composition of the modes at the beginning is low because there is
no coherent radiation at the beginning of the wiggler. Also, the composition
of the lowest order mode is 84.06\% at the suppression point and increases
to 93.86\% at the saturation point. And it is found that compositions
for modes $TEM_{20}$ and $TEM_{02}$ are 6.4\% and 5.6\% at the suppression
point, respectively, and they decrease to 2.6\% at the saturation
point for both modes. It is clear that mode $TEM_{00}$ is dominant
at the saturation point, which is reported in Table \ref{tab:Modal-composition-shot}.
At the saturation point, the composition of the lowest mode in SASE
FEL is larger than that of in the seeded FEL and higher modes are
less effective in SASE FEL compared to those in the seeded FEL. Because
in the SASE FEL, the power of radiation levels off completely at the
saturation point. However, the radiation power of seeded FEL has a
very small growth at the saturation point, therefor, higher number
of modes also contribute to the composition at the saturation point
compared to the higher modes of radiation in the SASE FEL.

\begin{figure}
\includegraphics[width=8cm,height=6cm]{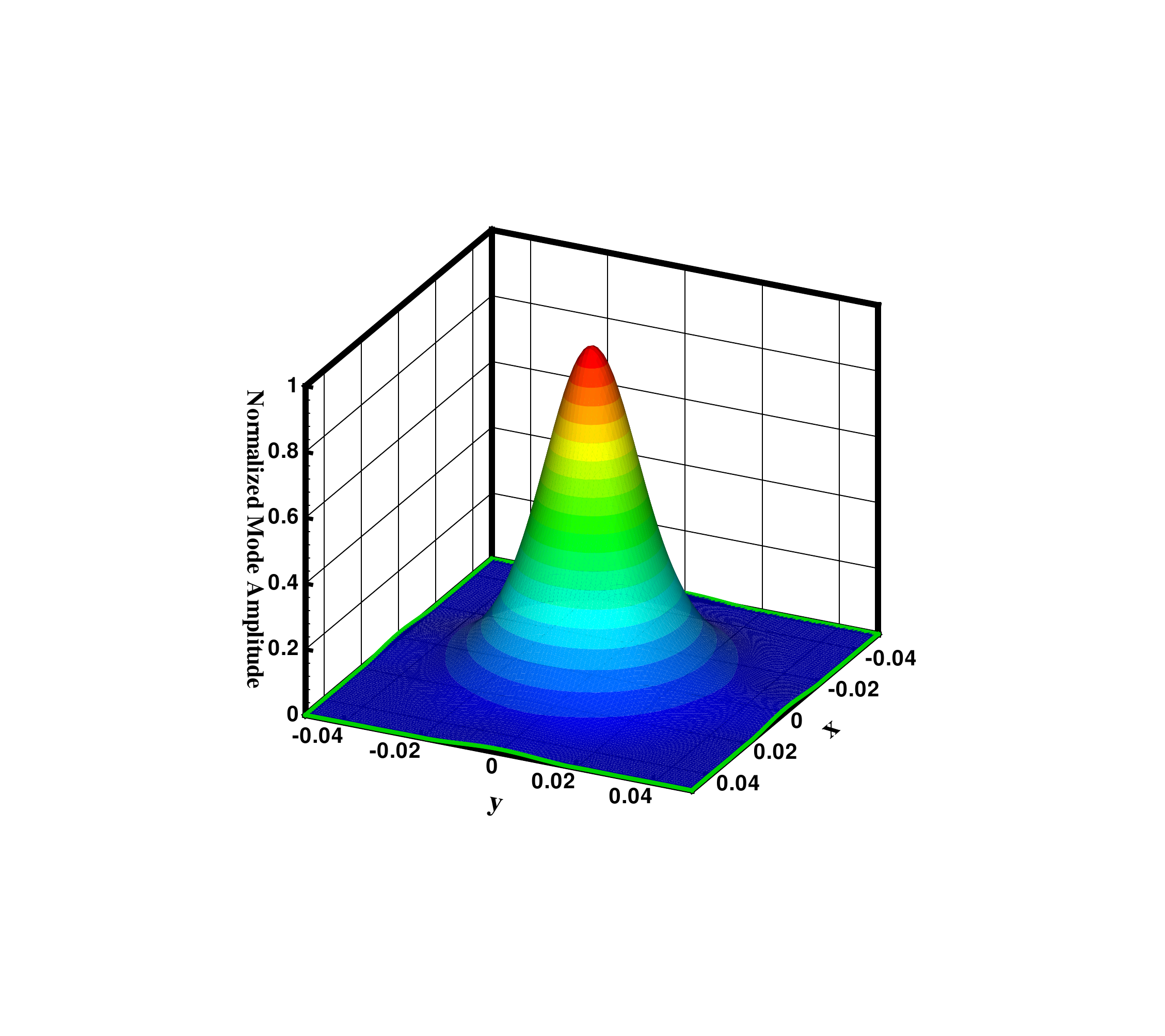}

\caption{Transverse mode pattern for the third harmonic at the saturation point
(for the third harmonic)\label{fig:amplitudeshot}}
\end{figure}

\begin{table}
\begin{tabular}{cccc}
\hline 
 & z=0 & z=11m & z=22.8\tabularnewline
\hline 
modes & \%composition & \%composition & \%composition\tabularnewline
\hline 
(0,0) & 17.6\% & 84.06\% & 93.86\%\tabularnewline
(0,1) & $\thickapprox$0.0\% & $\thickapprox$0.0\% & $\thickapprox0.0$\%\tabularnewline
(1,0) & $\thickapprox$0.0\% & $\thickapprox$0.0\% & $\thickapprox$0.0\%\tabularnewline
(0,2) & 12.9\% & 6.4\% & 2.6\%\tabularnewline
(2,0) & 12.9\% & 5.6\% & 2.6\%\tabularnewline
(2,2) & 8.2\% & 0.77\% & 0.21\%\tabularnewline
(0,4) & 12.4\% & 1.3\% & 0.3\%\tabularnewline
(4,0) & 12.4\% & 1.02\% & 0.3\%\tabularnewline
(2,4) & 7.8\% & 0.33\% & 0.05\%\tabularnewline
(4,2) & 7.8\% & 0.29\% & 0.05\%\tabularnewline
(4,4) & 7.7\% & 0.04\% & 0.016\%\tabularnewline
\end{tabular}

\caption{Modal composition of the third harmonic in different points of the
wiggler\label{tab:Modal-composition-shot}}
\end{table}
 The normalized intensity of the third harmonic near saturation point
is represented in Fig. \ref{fig:amplitudeshot}.

\section{Conclusion}

In this paper, we analyzed detuning of the fundamental to enhance
harmonic generation in x-ray FELs. It works well in the frame work
of the realistic 3D model of the FEL process by using a nonaveraged
simulations, which is named CYRUS 3D. In the absence of slippage (steady-state
simulation), the variation of radiation waists, curvatures, and amplitudes
for fundamental resonance and the third harmonic are studied. Transverse
mode evolution of the fundamental and the third harmonic are investigated
in more details. The radiation power of the third harmonic is larger
than that of the fundamental resonance in contrast to the nonlinear
harmonic generation. Also, the composition of significant modes of
the third harmonic is presented which shows that the lowest order
mode is dominant. We discussed shot noise in harmonic lasing FEL that
treats the start up of both the fundamental and harmonic generation.
The saturation power of the third harmonic is increased considerably
compared to that of the seeded FEL case while increase of the saturation
length is negligible in SASE FEL. Also $TEM_{00}$ mode is more effective
at the saturation point in SASE FEL compared to that in the seeded
FEL .
\begin{acknowledgments}
This work was supported by Institute for Research in Fundamental Sciences
(IPM) to allow us to use computational facilities. \end{acknowledgments}

\end{document}